\documentclass[twocolumn,epjc3]{svjour3} 
\usepackage[T1]{fontenc}

\smartqed  % flush right qed marks, e.g. at end of proof

\usepackage{amsmath,amsfonts,amssymb}
\usepackage{graphicx}
\usepackage{tgtermes}      
\usepackage{flushend}
\usepackage[numbers,sort&compress]{natbib}
\usepackage[colorlinks,citecolor=blue,urlcolor=blue,linkcolor=blue]{hyperref}

\usepackage{mathtools}

\journalname{Eur. Phys. J. C}

\newcommand{\ud}{\mathrm{d}}
\newcommand{\ve}{\varepsilon}

\begin{document}

\title{The non-minimal coupling constant and the primordial de Sitter state}

\author{Orest Hrycyna\thanksref{e1,addr1}}

\institute{Theoretical Physics Division, National Centre for Nuclear
Research, Ludwika Pasteura 7, 02-093 Warszawa, Poland\label{addr1}}
\thankstext{e1}{e-mail: orest.hrycyna@ncbj.gov.pl}

\date{Received: \today}

\maketitle

\begin{abstract}
Dynamical systems methods are used to investigate dynamics of a flat Friedmann--Robertson--Walker cosmological model with the non-minimally coupled scalar field and a potential function. Performed analysis distinguishes the value of non-minimal coupling constant parameter $\xi=\frac{3}{16}$, which is the conformal coupling in five dimensional theory of gravity. It is shown that for a monomial potential functions at infinite values of the scalar field there exist generic de Sitter and Einstein-de Sitter states. The de Sitter state is unstable with respect to expansion of the Universe for potential functions which do not change faster than linearly. This leads to a generic cosmological evolution without the initial singularity.
\end{abstract}


\section{Introduction}
The theoretical cosmology explores past and future evolution of the Universe. Our methodological tools are based on models derived from general relativity. The unexpected discovery of accelerated expansion of the Universe \cite{Riess:1998cb, Perlmutter:1998np} has provoked growth of interest in dynamical dark energy models \cite{Copeland:2006wr, Bahamonde:2017ize}. The simples model cosmological model with a scalar field and a potential function can serve as prototype to describe an accelerated expansion of the Universe \cite{Ratra:1987rm, Wetterich:1987fm}. This is so-called a quintessence idea. 

A scalar field matter sector of a cosmological theory with an non-minimal coupling term $-\xi R\phi^{2}$ between the gravity and the scalar field where $\xi$ is the dimensionless coupling constant \cite{Chernikov:1968zm, Callan:1970ze, Birrell:1979ip} give rise to the simplest extension of the scalar field Lagrangian. There are various motivations for such term. The most general one is that the general relativity has a methodological status of an effective theory and such term naturally emerge in its expansion \cite{Donoghue:1994dn}. The 
non-minimal coupling between the scalar curvature and the scalar field appears as a result of quantum corrections to the scalar field in curved space and the renormalisation procedure also give rise to such term \cite{Allen:1983dg, Ishikawa:1983kz, Birrell:1984ix, Parker:book}. 
The non-minimal coupling is also interesting in the context of 
superstring theory \cite{Maeda:1985bq} and induced gravity \cite{Accetta:1985du}. 

While the simplest inflationary model with a minimally coupled scalar field and a quadratic potential function is no longer favoured by the observational data \cite{Planck:2013jfk, Martin:2013nzq, Ade:2015lrj, Kobayashi:2011nu} there is a need to extend this paradigm further. 
From the theoretical point of view and an effective theory approach the coupling constant becomes a free parameter in the model and should be obtained from some general considerations \cite{Atkins:2010eq, Atkins:2010re} or from a more fundamental theory.
Taking a pragmatic approach its value should be estimated from the observational data \cite{Luo:2005ra, Nozari:2007eq, Szydlowski:2008zza, Atkins:2012yn, Hrycyna:2015vvs}.

The non-minimally coupled scalar field cosmology was investigated by many authors in the connection with an inflationary epoch as well as a description of the current accelerated expansion of the universe \cite{Spokoiny:1984bd, Ford:1986sy, Salopek:1988qh, Amendola:1990nn, Fakir:1992cg, Barvinsky:1994hx, Faraoni:1996rf, Barvinsky:1998rn, Barvinsky:2008ia, Setare:2008mb, Setare:2008pc, Uzan:1999ch, Chiba:1999wt, Amendola:1999qq, Holden:1999hm, Bartolo:1999sq, Boisseau:2000pr, Gannouji:2006jm, Carloni:2007eu, Bezrukov:2007ep, Kamenshchik:1995ib,Hrycyna:2007gd, Hrycyna:2008gk, Hrycyna:2009zj, Hrycyna:2010yv, Hrycyna:2015eta}. In the standard model of particle physics a non-minimally coupled Higgs field plays also important role \cite{DeSimone:2008ei, Bezrukov:2008ej, Barvinsky:2009fy, Clark:2009dc}.

In current investigations in modern theoretical cosmology dynamical systems methods constitute very strong and effective tools 
\cite{Belinskii:1985, Belinsky:1985zd, Belinskii:1987, Wainwright:book}. A cosmological evolution of a universe is represented by trajectories in a space of all states of a model called a phase space. In the recent paper \cite{Humieja:2019ywy} the authors used interesting tools of bifurcation theory in dynamical systems of cosmological origin and found bifurcation values of parameters of the investigated models where qualitative dynamics changes. 

We start with the total action integral for the gravitational theory under considerations
\begin{equation}
\label{eq:action}
S =S_{g} + S_{\phi}\,,
\end{equation}
where the gravitational part of the theory is given by the standard Einstein-Hilbert action integral
\begin{equation}
S_{g} = \frac{1}{2\kappa^{2}}\int\ud^{4}x\sqrt{-g}\,R \,,
\end{equation}
where $\kappa^{2}=8\pi G$, and $G$ is the Newton constant. The matter part of the theory is described by the scalar field action with additional term describing direct interaction between the Ricci scalar and the scalar field 
\begin{equation}
S_{\phi}=- \frac{1}{2}\int\ud^{4}x\sqrt{-g}\Big(\ve\nabla^{\alpha}\phi\,\nabla_{\alpha}\phi + \ve\xi R \phi^{2} + 2U(\phi)\Big)\,,
\end{equation}
where $\ve=+1$ and $\ve=-1$ correspond to the canonical and phantom scalar field, respectively, and $\xi$ is the non-minimal coupling constant between the scalar field and gravity.

The field equations for the theory are
\begin{equation}
R_{\mu\nu}-\frac{1}{2}g_{\mu\nu}R=\kappa^{2}\,T^{(\phi)}_{\mu\nu}\,,
\end{equation}
where the energy-momentum tensor for the non-minimally coupled scalar field is given by
\begin{equation}
\label{eq:energy_mom}
\begin{split}
T^{(\phi)}_{\mu\nu}= &\,\,\ve \nabla_{\mu}\phi\nabla_{\nu}\phi -
\ve\frac{1}{2}g_{\mu\nu}\nabla^{\alpha}\phi\nabla_{\alpha}\phi -
U(\phi)g_{\mu\nu}  \\ & + \ve\xi\Big(R_{\mu\nu}-\frac{1}{2}g_{\mu\nu}R\Big)\phi^{2} +
\ve\xi\Big(g_{\mu\nu}\Box\phi^{2}-\nabla_{\mu}\nabla_{\nu}\phi^{2}\Big)\,.
\end{split}
\end{equation}
Finally, the dynamical equation for the scalar field we obtain from the variation $\delta S_{\phi}/\delta\phi=0$  
\begin{equation}
\Box\phi-\xi R\phi-\ve \,U'(\phi)=0\,.
\end{equation}

In the present paper we investigate exclusively the original Jordan frame formulation of the theory leaving aside the problem of the physical (in)equivalence between the Jordan frame and the Einstein frame \cite{Faraoni:1999hp, Kamenshchik:2014waa, Bahamonde:2017kbs, Calmet:2017voc}.

Working with the spatially flat Friedmann--Robertson--Walker metric
$$
\ud s^{2}=-\ud t^{2}+a^{2}(t)\Big(\ud x^{2}+\ud y^{2}+\ud z^{2}\Big)\,,
$$
and assuming homogeneous scalar field $\phi=\phi(t)$
we obtain the energy conservation condition 
\begin{equation}
\label{eq:constr}
\frac{3}{\kappa^{2}}H^{2}=\rho_{\phi}=\ve\frac{1}{2}\dot{\phi}^{2}+U(\phi)+\ve3\xi H^{2}\phi^{2}+\ve6\xi H \phi\dot{\phi}\,,
\end{equation}
the acceleration equation
\begin{equation}
\label{eq:accel}
\dot{H}=-2H^{2}+\frac{\kappa^{2}}{6}\frac{-\ve(1-6\xi)\dot{\phi}^{2}+4U(\phi)-6\xi\phi U'(\phi)}{1-\ve\xi(1-6\xi)\kappa^{2}\phi^{2}}\,,
\end{equation}
and the equation of motion for the scalar field
\begin{equation}
\ddot{\phi}+3H\dot{\phi}+6\xi\big(\dot{H}+2H^{2}\big)\phi + \ve U'(\phi)=0\,,
\end{equation}
where a dot indicates differentiation with respect to cosmological time $t$.

In what follows we introduce the following dimensionless dynamical variables 
\cite{Hrycyna:2015eta}
\begin{equation}
\begin{split}
& x\equiv\frac{\kappa\dot{\phi}}{\sqrt{6}H_{0}}\,,\quad y\equiv\frac{\kappa\sqrt{U(\phi)}}{\sqrt{3}H_{0}}\,,\quad z\equiv\frac{\kappa}{\sqrt{6}}\frac{H}{H_{0}}\phi\,, \\ & h\equiv\frac{H}{H_{0}}\,,\quad \lambda\equiv-\phi\frac{U'(\phi)}{U(\phi)}\,.
\end{split}
\end{equation}
Then, the energy conservation condition \eqref{eq:constr} is
\begin{equation}
\label{eq:constr_new}
h^{2}=y^{2}+\ve(1-6\xi)x^{2}+\ve6\xi(x+z)^{2}\,,
\end{equation}
and the acceleration equation \eqref{eq:accel} expressed in the new phase space variables
\begin{equation}
\frac{\dot{H}}{H^{2}} = -2 + \frac{-\ve(1-6\xi)x^{2}+y^{2}(2+3\xi\lambda)}{h^{2}-\ve6\xi(1-6\xi)z^{2}}\,.
\end{equation}

The full dynamical phase space of the model is five dimensional and the autonomous dynamical system describing evolution of the model is in the following form
\begin{equation}
\label{eq:dyn_sys_full}
\begin{split}
\frac{\ud x}{\ud \ln{a}} & = -3x-6\xi z\bigg(\frac{\dot{H}}{H^{2}}+2\bigg) + \ve\frac{1}{2}\lambda y^{2}\frac{1}{z}\,,\\
\frac{\ud y}{\ud \ln{a}} & =-\frac{1}{2}\lambda y \frac{x}{z}\,,\\
\frac{\ud z}{\ud \ln{a}} & = x + z\frac{\dot{H}}{H^{2}}\,,\\
\frac{\ud h}{\ud \ln{a}} & = h\frac{\dot{H}}{H^{2}}\,,\\
\frac{\ud \lambda}{\ud \ln{a}} & = \frac{x}{z}\lambda\Big(1-\lambda\big(\Gamma-1\big)\Big)\,,
\end{split}
\end{equation}
where
\begin{equation}
\Gamma=\frac{U''(\phi)U(\phi)}{U'(\phi)^{2}}\,,
\end{equation}
and a prime denotes differentiation with respect to the scalar field and a ''time'' parameter along the phase space curves is natural logarithm of the scale factor.

Now, using the energy conservation condition \eqref{eq:constr_new} one can eliminate one of the phase space variables, say $h$ or $y$, hence one obtains a $4$-dimensional dynamical system. Assuming that $\Gamma=\Gamma(\lambda)$ and using the second and the last equation of the system one obtains the following differential equation
$$
\frac{\ud y}{\ud \lambda}= -\frac{1}{2}\frac{y}{\lambda\Big(1-\lambda\big(\Gamma(\lambda)-1\big)\Big)}\,,
$$
which can be solved for some generic function $\Gamma(\lambda)$ and one can further reduce dynamical system \eqref{eq:dyn_sys_full} to a $3$-dimensional phase space \cite{Hrycyna:2010yv}.

\section{An asymptotically monomial potential function}

In this section we will investigate behaviour of the model in the limit $\phi\to\infty$ and we assume that $\lambda\to\text{const.}$ in this limit. One can observe that within this assumptions we have a so-called plateau-like scalar field potential functions where $\frac{U'(\phi)}{U(\phi)}\to0$  as $\phi\to\infty$ and, moreover, not only potential functions with $U(\phi)\to\text{const.}$ as $\phi\to\infty$ but all possible potential functions with a monomial asymptotic behaviour. In other words we assume that some general scalar field potential function has an asymptotic in the form of monomial potential function $U(\phi)\propto \phi^{\alpha}$ as value of the scalar field  $\phi>>\phi^{*}$ is much greater than some given value. This assumption can be motivated by the recent observational data \cite{Planck:2013jfk, Martin:2013nzq, Ade:2015lrj, Planck2018:X} which favour flat inflationary scalar field potential and also by a pragmatic approach in order to constrain an unknown function of the theory \cite{Alho:2015cza}.

Using the energy conservation condition \eqref{eq:constr_new} we eliminate the variable $h^{2}$ from the dynamical system \eqref{eq:dyn_sys_full} and with an asymptotically monomial potential function where $\lambda=\text{const.}$ we obtain the following set of equations
\begin{equation}
\label{eq:dynsys_arb}
\begin{split}
\frac{\ud x}{\ud \ln{a}} & = -3 x - 6\xi z\bigg(\frac{\dot{H}}{H^{2}}+2\bigg)+\ve\frac{1}{2}\lambda y^{2}\frac{1}{z}\,,\\
\frac{\ud y}{\ud \ln{a}} & = -\frac{1}{2}\lambda y \frac{x}{z}\,,\\
\frac{\ud z}{\ud \ln{a}} & = x + z\frac{\dot{H}}{H^{2}}\,,
\end{split}
\end{equation}
where the acceleration equation is now
\begin{equation}
\frac{\dot{H}}{H^{2}} = -2 + \frac{-\ve(1-6\xi)x^{2}+(2+3\xi\lambda)y^{2}}{y^{2}+\ve(x+6\xi z)^{2}}\,,
\end{equation}
and the energy conservation condition is given by \eqref{eq:constr_new} which was used to eliminate the $h$ variable from the dynamical system \eqref{eq:dyn_sys_full}.

Note that the system \eqref{eq:dynsys_arb} is homogeneous and linear in dynamical phase space variables. In this case, dimension of dynamical system can be reduced using new appropriate projective coordinates \cite{Perko:book, Wiggins:book}.

Introducing the following projective coordinates
\begin{equation}
\label{eq:proj_coor}
u\equiv\frac{x}{z}\,,\,\, \bar{v}= v^{2}\equiv\frac{y^{2}}{z^{2}}\,,\,\, \bar{w}=w^{2}\equiv\frac{1}{z^{2}}\,,
\end{equation}
we obtain the dynamical system in the new variables
\begin{equation}
\label{eq:dynsys_arb_inf}
\begin{split}
\frac{\ud u}{\ud\eta} & =  \left(-u(u+1)+\ve\frac{1}{2}\lambda \bar{v}\right)
\big(\bar{v}+\ve(u+6\xi)^{2}\big) \\ & \hspace{5mm} +(u+6\xi)\big(\ve(1-6\xi)u^{2}-(2+3\xi\lambda)\bar{v}\big)\,,\\
\frac{\ud \bar{v}}{\ud\eta} & =  2\bar{v}\bigg( \left(2-\frac{1}{2}(2+\lambda)u\right)\big(\bar{v}+\ve(u+6\xi)^{2}\big) \\ & \hspace{1cm} +\ve(1-6\xi)u^{2}-(2+3\xi\lambda)\bar{v}\bigg)\,,\\
\frac{\ud \bar{w}}{\ud\eta} & =  2\bar{w}\Big((2-u)\big(\bar{v}+\ve(u+6\xi)^{2}\big) \\ & \hspace{1cm} +\ve(1-6\xi)u^{2}-(2+3\xi\lambda)\bar{v}\Big)\,,
\end{split}
\end{equation}
where the time parameter
\begin{equation}
\frac{\ud}{\ud\eta}=\big(\bar{v}+\ve(u+6\xi)^{2}\big)\frac{\ud}{\ud\ln{a}}\,,
\end{equation}
was introduced in order to put all the functions on the right hand side into a polynomial form.  The acceleration equation now reads
\begin{equation}
\label{eq:accel_arb_inf}
\frac{\dot{H}}{H^{2}} = -2+\frac{-\ve(1-6\xi)u^{2}+(2+3\xi\lambda)\bar{v}}{\bar{v}+\ve(u+6\xi)^{2}}\,,
\end{equation}
and the energy conservation condition is
\begin{equation}
\label{eq:hubble_arb_inf}
h^{2}=\frac{\bar{v}+\ve(1-6\xi)u^{2}+\ve6\xi(u+1)^{2}}{\bar{w}}\,.
\end{equation}
Note that the first two equations of the system \eqref{eq:dynsys_arb_inf} are now decoupled from the third giving rise to effective two dimensional dynamical system describing the evolution of the model under considerations. Using linearised solutions to the decoupled two dimensional part of the system we can solve the third equation in the exact form.

\begin{table*}
\caption{The critical points of the decoupled two dimensional part of the system \eqref{eq:dynsys_arb_inf} and corresponding values of the acceleration equation \eqref{eq:accel_arb_inf}.}
\label{tab:1}
\renewcommand{\arraystretch}{1.5}
	\centering
	\begin{tabular}{|c|l|l|c|}
		\hline
		& $u^{*}$ & $\bar{v}^{*}$ & $\frac{\dot{H}}{H^{2}}\big|^{*}$ \\
		\hline
		1. & $0$ & $0$ & $-2$ \\
		2. & $-6\xi$ & $0$ & $\pm\infty$  \\
		3. & $-6\xi\pm\sqrt{-6\xi(1-6\xi)}$ & $0$ & $-2+\frac{(u^{*})^{2}}{6\xi}$ \\
		4. & $\frac{-2\xi(2+3\lambda\xi)\pm2\xi\sqrt{-(1-6\xi)(2+3\lambda\xi)}}{1-(2-\lambda)\xi}$ &
		$-\ve(u^{*}+6\xi)^{2}$ & $0$ \\
		5. & $0$ & $\ve\frac{24\xi}{\lambda}$ & $0$ \\
		6. & $-\frac{(4+\lambda)\xi}{1-(2-\lambda)\xi}$ & $-\ve\frac{(1-6\xi)\xi(6-(2-\lambda)(10+\lambda)\xi)}{(1-(2-\lambda)\xi)^{2}}$ & $\frac{1}{2}\frac{(2+\lambda)(4+\lambda)\xi}{1-(2-\lambda)\xi}$  \\
		\hline
	\end{tabular}
\end{table*}

In Table \ref{tab:1} we have gathered the critical points of the decoupled $2$-dimensional dynamical system \eqref{eq:dynsys_arb_inf} in variables $(u,\bar{v})$ together with the corresponding values of the acceleration equations \eqref{eq:accel_arb_inf} calculated at those states. The first critical point corresponds to a radiation-like expansion of the universe and it is in the form of a saddle type critical point. From the physical point of view very interesting is the last critical point. The acceleration equation vanishes for two values of $\lambda$ parameter, namely, for $\lambda=-2$ and $\lambda=-4$, which corresponds to a quadratic and a quartic asymptotic form of the potential function \cite{Dutta:2020uha}. Vanishing of the acceleration equation could possibly indicate to the de Sitter type of evolution and more thoroughly analysis is required. Due to their special importance in particle physics and cosmology, an asymptotic form of the scalar field potential function with $\lambda=-2$ and $\lambda=-4$ will be discussed elsewhere. The critical point $5$ with the vanishing acceleration equation corresponds to de Sitter state which can be made asymptotically stable for some set of the model parameters. Note that this state is defined for an arbitrary value of the $\bar{w}$ variable and is described by a non-hyperbolic critical point (one of eigenvalues of the linearisation matrix calculated at this point vanishes) and in order to find linearised solutions in the vicinity of this de Sitter state first we need to solve decoupled subsystem \eqref{eq:dynsys_arb_inf} and then solve the third equation. Dynamical analysis of these states is beyond scope of the present work.

Finally, let us note that the critical points $3$ give rise to two interesting asymptotic states for the special value of the non-minimal coupling parameter $\xi=\frac{3}{16}$. In the first case we obtain $\frac{\dot{H}}{H^{2}}\big|^{*}=-\frac{3}{2}$ which corresponds to the Einstein--de Sitter type of evolution, while in the second case we have $\frac{\dot{H}}{H^{2}}\big|^{*}=0$ which correspond to the de Sitter expansion. In what follows we concentrate our investigations 
on those states with $\xi=\frac{3}{16}$ which is the value of the non-minimal coupling constant for a conformally coupled scalar field in a $5$-dimensional theory of gravity.

\section{$\xi=\frac{3}{16}$ and instability of the initial de Sitter state}

Dynamical system \eqref{eq:dynsys_arb} in projective coordinates \eqref{eq:proj_coor} and for $\xi=\frac{3}{16}$ takes the following form
\begin{equation}
\label{eq:dynsys_reduced}
\begin{split}
\frac{\ud u}{\ud\ln{a}} & = - u(u+1)+\ve\frac{1}{2}\lambda \bar{v} \\ & \hspace{5mm} -\left(u+\frac{9}{8}\right)\frac{\ve\frac{1}{8}u^{2}+\left(2+\frac{9}{16}\lambda\right)\bar{v}}{\bar{v}+\ve\left(u+\frac{9}{8}\right)^{2}}\,,\\
\frac{\ud \bar{v}}{\ud\ln{a}} & = 2\bar{v}\left( 2-\frac{1}{2}(2+\lambda)u  -\frac{\ve\frac{1}{8}u^{2}+\left(2+\frac{9}{16}\lambda\right)\bar{v}}{\bar{v}+\ve\left(u+\frac{9}{8}\right)^{2}}\right)\,,\\
\frac{\ud \bar{w}}{\ud\ln{a}} & = 2\bar{w}\left(2-u-\frac{\ve\frac{1}{8}u^{2}+\left(2+\frac{9}{16}\lambda\right)\bar{v}}{\bar{v}+\ve\left(u+\frac{9}{8}\right)^{2}}\right)\,,
\end{split}
\end{equation}
where the time parameter along the phase space trajectories in the natural logarithm of the scale factor.

The first critical point under considerations 
$$
\left(u^{*}=-\frac{3}{4}\,,\bar{v}^{*}=0\,,\bar{w}^{*}=0\right)
$$
with the acceleration equation $\frac{\dot{H}}{H^{2}}\big|^{*}=-\frac{3}{2}$ has the following linearised solutions
\begin{equation}
\label{eq:lin_sol_eds}
\begin{split}
u(a) & = u^{*}+\left(\Delta u +\ve\frac{4}{3}\Delta\bar{v}\right)\left(\frac{a}{a^{(i)}}\right)^{\frac{3}{2}} \\ & \hspace{9mm}-\ve\frac{4}{3}\Delta\bar{v}\left(\frac{a}{a^{(i)}}\right)^{\frac{3}{4}(6+\lambda)}\,,\\
\bar{v}(a) & = \Delta\bar{v}\left(\frac{a}{a^{(i)}}\right)^{\frac{3}{4}(6+\lambda)}\,,\\
\bar{w}(a) & = \Delta\bar{w}\left(\frac{a}{a^{(i)}}\right)^{\frac{9}{2}}\,,
\end{split}
\end{equation}
where $\Delta u = u^{(i)} - u^{*}$, $\Delta\bar{v}=\bar{v}^{(i)}$ and $\Delta\bar{w}=\bar{w}^{(i)}$ are the initial conditions for the phase space variables and $a^{(i)}$ is the initial value of the scale factor in the vicinity of the critical point. The asymptotic state is unstable with respect to the expansion of universe for $6+\lambda>0$, otherwise it corresponds to a saddle type critical point. 

Taking those linearised solution one can calculate the acceleration equation \eqref{eq:accel_arb_inf} up to linear terms in initial conditions
\begin{equation}
\begin{split}
\frac{\dot{H}}{H^{2}} & \approx - \frac{3}{2} - 4\left(\Delta u +\ve\frac{4}{3}\Delta\bar{v}\right)\left(\frac{a}{a^{(i)}}\right)^{\frac{3}{2}} \\ & \hspace{7mm} +\ve 4(4+\lambda)\Delta\bar{v}\left(\frac{a}{a^{(i)}}\right)^{\frac{3}{4}(6+\lambda)}\,,
\end{split}
\end{equation}
which for $a\to0$ converges to $\frac{\dot{H}}{H^{2}}\big|^{*}=-\frac{3}{2}$. The Hubble function \eqref{eq:hubble_arb_inf} up to linear order in the initial conditions is in following form
%\begin{widetext}
\begin{equation}
\begin{split}
\left(\frac{H(a)}{H(a_{0})}\right)^{2}\approx\frac{1}{\Delta\bar{w}}\Bigg(&\left(\Delta\bar{v}+\ve\frac{3}{4}\Delta u\right)\left(\frac{a}{a^{(i)}}\right)^{-3}\\ & + \ve\left(\Delta u +\ve\frac{4}{3}\Delta\bar{v}\right)^{2}\left(\frac{a}{a^{(i)}}\right)^{-\frac{3}{2}}  \\
& -\frac{8}{3}\Delta\bar{v}\left(\Delta u +\ve\frac{4}{3}\Delta\bar{v}\right)\left(\frac{a}{a^{(i)}}\right)^{\frac{3}{4}(2+\lambda)}  \\ & + \ve\frac{16}{9}\Delta\bar{v}^{2}\left(\frac{a}{a^{(i)}}\right)^{\frac{3}{2}(3+\lambda)}\Bigg)\,,
\end{split}
\end{equation}
%\end{widetext}
\noindent
where the first term dominates for a given set of initial conditions. As $a\to0$ we need to impose $5+\lambda>0$ in order to ensure that the last two terms do not become large. Then the Hubble function can be approximated as
\begin{equation}
\left(\frac{H(a)}{H(a_{0})}\right)^{2} \approx \frac{\Delta\bar{v}+\ve\frac{3}{4}\Delta u}{\Delta\bar{w}}\left(\frac{a}{a^{(i)}}\right)^{-3}\,,
\end{equation}
which corresponds exactly to the Einstein--de Sitter type of evolution and takes place in the physical region of the phase space for $\Delta\bar{v}+\ve\frac{3}{4}\Delta u>0$.

The initial value of the Hubble function at $a=a^{(i)}$ can be expressed as
\begin{equation}
\left(\frac{H(a^{(i)})}{H(a_{0})}\right)^{2} \approx  \frac{\Delta\bar{v}+\ve\frac{3}{4}\Delta u}{\Delta\bar{w}} +\ve\frac{\Delta u^{2}}{\Delta\bar{w}}>0\,,
\end{equation}
and we obtain evolutionary equation for the Hubble function in the vicinity of the critical point corresponding to the Einstein--de Sitter type of solution
\begin{equation}
\left(\frac{H(a)}{H(a_{0})}\right)^{2} \approx \left(\left(\frac{H(a^{(i)})}{H(a_{0})}\right)^{2}- \ve\frac{\Delta u^{2}}{\Delta\bar{w}}\right)\left(\frac{a}{a^{(i)}}\right)^{-3}\,.
\end{equation}

The second critical point under considerations $$\left(u^{*}=-\frac{3}{2}\,,\bar{v}^{*}=0\,,\bar{w}^{*}=0\right)$$ with the vanishing acceleration equation $\frac{\dot{H}}{H^{2}}\big|^{*}=0$ has the following linearised solutions 
\begin{equation}
\label{eq:lin_sol_ds}
\begin{split}
u(a) & = u^{*}+\left(\Delta u -\ve\frac{4}{3}\Delta\bar{v}\right)\left(\frac{a}{a^{(i)}}\right)^{3} \\ & \hspace{8.5mm} +\ve\frac{4}{3}\Delta\bar{v}\left(\frac{a}{a^{(i)}}\right)^{\frac{3}{2}(2+\lambda)}\,,\\
\bar{v}(a) & = \Delta\bar{v}\left(\frac{a}{a^{(i)}}\right)^{\frac{3}{2}(2+\lambda)}\,,\\
\bar{w}(a) & = \Delta\bar{w}\left(\frac{a}{a^{(i)}}\right)^{3}\,,
\end{split}
\end{equation}
where $\Delta u = u^{(i)} - u^{*}$, $\Delta\bar{v}=\bar{v}^{(i)}$ and $\Delta\bar{w}=\bar{w}^{(i)}$ are the initial conditions for the phase space variables and $a^{(i)}$ is the initial value of the scale factor in the vicinity of the asymptotic state. Dynamical character of this critical point corresponds to an unstable node with respect to the expansion of universe for $2+\lambda>0$, otherwise, this state is represented by a saddle type critical point.

The linearised solutions can be used to obtain the acceleration equation \eqref{eq:accel_arb_inf} in the vicinity of the state up to linear terms in initial conditions
\begin{equation}
\begin{split}
\frac{\dot{H}}{H^{2}} & \approx 8\left(\Delta u -\ve\frac{4}{3}\Delta\bar{v}\right)\left(\frac{a}{a^{(i)}}\right)^{3} \\ & \hspace{5mm} +\ve \frac{4}{3}(8+3\lambda)\Delta\bar{v}\left(\frac{a}{a^{(i)}}\right)^{\frac{3}{2}(2+\lambda)}\,,
\end{split}
\end{equation}
and the Hubble function \eqref{eq:hubble_arb_inf} as
%\begin{widetext}
\begin{equation}
\begin{split}
\left(\frac{H(a)}{H(a_{0})}\right)^{2}\approx \frac{1}{\Delta\bar{w}}\Bigg(&
\left(\Delta\bar{v}-\ve\frac{3}{4}\Delta u\right) \\ & +\ve\left(\Delta u -\ve\frac{4}{3}\Delta\bar{v}\right)^{2}\left(\frac{a}{a^{(i)}}\right)^{3} \\ 
&+ \frac{8}{3}\Delta\bar{v}\left(\Delta u -\ve\frac{4}{3}\Delta\bar{v}\right)\left(\frac{a}{a^{(i)}}\right)^{\frac{3}{2}(2+\lambda)} \\ & + \ve\frac{16}{9}\Delta\bar{v}^{2}\left(\frac{a}{a^{(i)}}\right)^{3(1+\lambda)}\Bigg)\,.
\end{split}
\end{equation}
%\end{widetext}
\noindent
We can note that the first term in this expression is zeroth order in the initial conditions and dominates during the evolution of the model while the remaining are first order terms in the initial conditions. Thus, taking the limit $a\to0$ and for $1+\lambda>0$ we obtain the Hubble function value at the critical point under considerations
\begin{equation}
\left(\frac{H(0)}{H(a_{0})}\right)^{2}\approx\frac{\Delta\bar{v}-\ve\frac{3}{4}\Delta u}{\Delta\bar{w}} = \textrm{const.} >0\,.
\end{equation}
The energy density of the initial de Sitter state is finite and depends on the initial conditions of trajectories leading to this state. Since $\Delta\bar{w}>0$ the only conditions for the state in the physical region of the phase space is $\Delta\bar{v}-\ve\frac{3}{4}\Delta u>0$. There is open and dense set of initial conditions in the vicinity of this state where energy density of the initial de Sitter state much larger than the present energy density.

Next, using the Hubble function \eqref{eq:hubble_arb_inf} we can relate the energy density at the initial conditions for the linearised solutions for $a=a^{(i)}$ with the energy density at the de Sitter state with $a=0$
\begin{equation}
\begin{split}
\left(\frac{H(a^{(i)})}{H(a_{0})}\right)^{2} & \approx \frac{\Delta\bar{v}-\ve\frac{3}{4}\Delta u}{\Delta\bar{w}}+\ve\frac{\Delta u^{2}}{\Delta\bar{w}} \\ & \approx \left(\frac{H(0)}{H(a_{0})}\right)^{2}+\ve\frac{\Delta u^{2}}{\Delta\bar{w}} >0\,,
\end{split}
\end{equation}
and since $\Delta\bar{w}$ is always positive, the energy density at $a=a^{(i)}>0$ can be larger than the energy density of the de Sitter state at $a=0$.

Now we are ready to show that the two critical points under considerations are indeed located at the infinite values of the scalar field $\phi$. From the energy conservation condition \eqref{eq:hubble_arb_inf} we obtain that $$\frac{6}{\kappa^{2}}\frac{1}{\phi^{2}}=  h^{2}\bar{w} =  \bar{v}+\ve\left(u+\frac{3}{4}\right)\left(u+\frac{3}{2}\right)$$ 
and using the linearised solutions \eqref{eq:lin_sol_eds} and \eqref{eq:lin_sol_ds} this quantity vanishes at the asymptotic states when $a\to0$ which indicates that $\phi\to\infty$ there. Obviously we can use linearised solutions for the variable $u$ and find linearised solutions for the field $\phi(a)$. 

It is worth noticing some analogies between asymptotic states under considerations and fast-roll (or rapid-roll) inflationary states \cite{Linde:2001ae,Kofman:2007tr,Chiba:2008ia}. The dynamical variable $u$ is defined as $u=\frac{\dot{\phi}}{H\phi}$ and the condition for the rapid-roll inflation for conformal coupling is given by $\dot{\phi}=- H \phi$ \cite{Kofman:2007tr} which for arbitrary coupling can be generalised as $$ \dot{\phi}=- 6\xi H \phi. $$ From Table \ref{tab:1} we have that in our case $$\dot{\phi}=\left(-6\xi\pm\sqrt{-6\xi(1-6\xi)}\right) H \phi$$
and for $\xi=\frac{3}{16}$ and the de Sitter state we have $$ \dot{\phi}=-\frac{3}{2}H\phi.$$
Since at the asymptotic de Sitter state the Hubble function $H\to\text{const.}$ and the scalar field $\phi\to\infty$ it gives that $\dot{\phi}\to-\infty$ and the scalar field changes infinitely fast with respect to the cosmological time.

\begin{figure*}
\centering
\begin{tabular}{cc}
\includegraphics[scale=0.75]{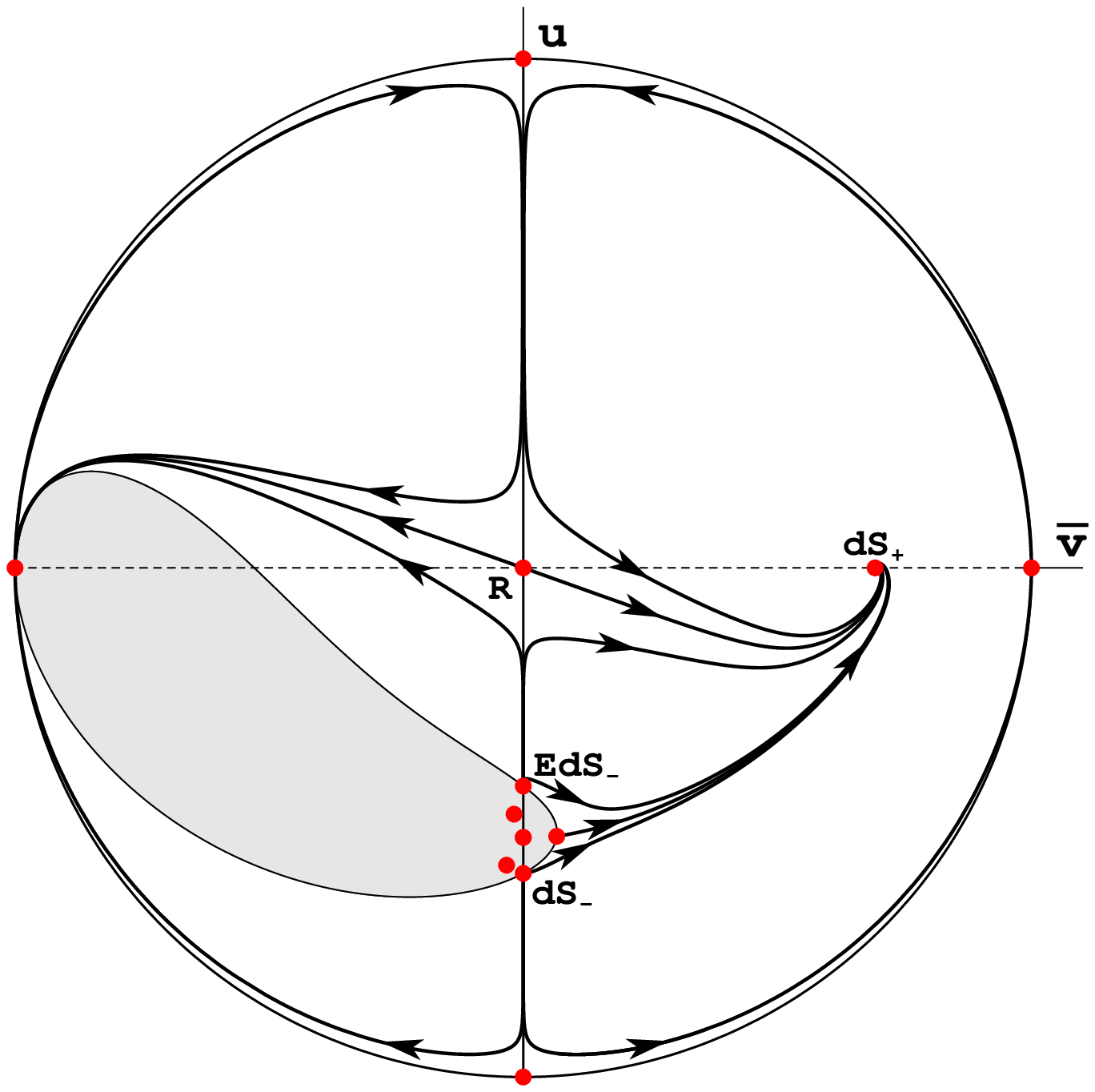} & \\
\includegraphics[scale=0.75]{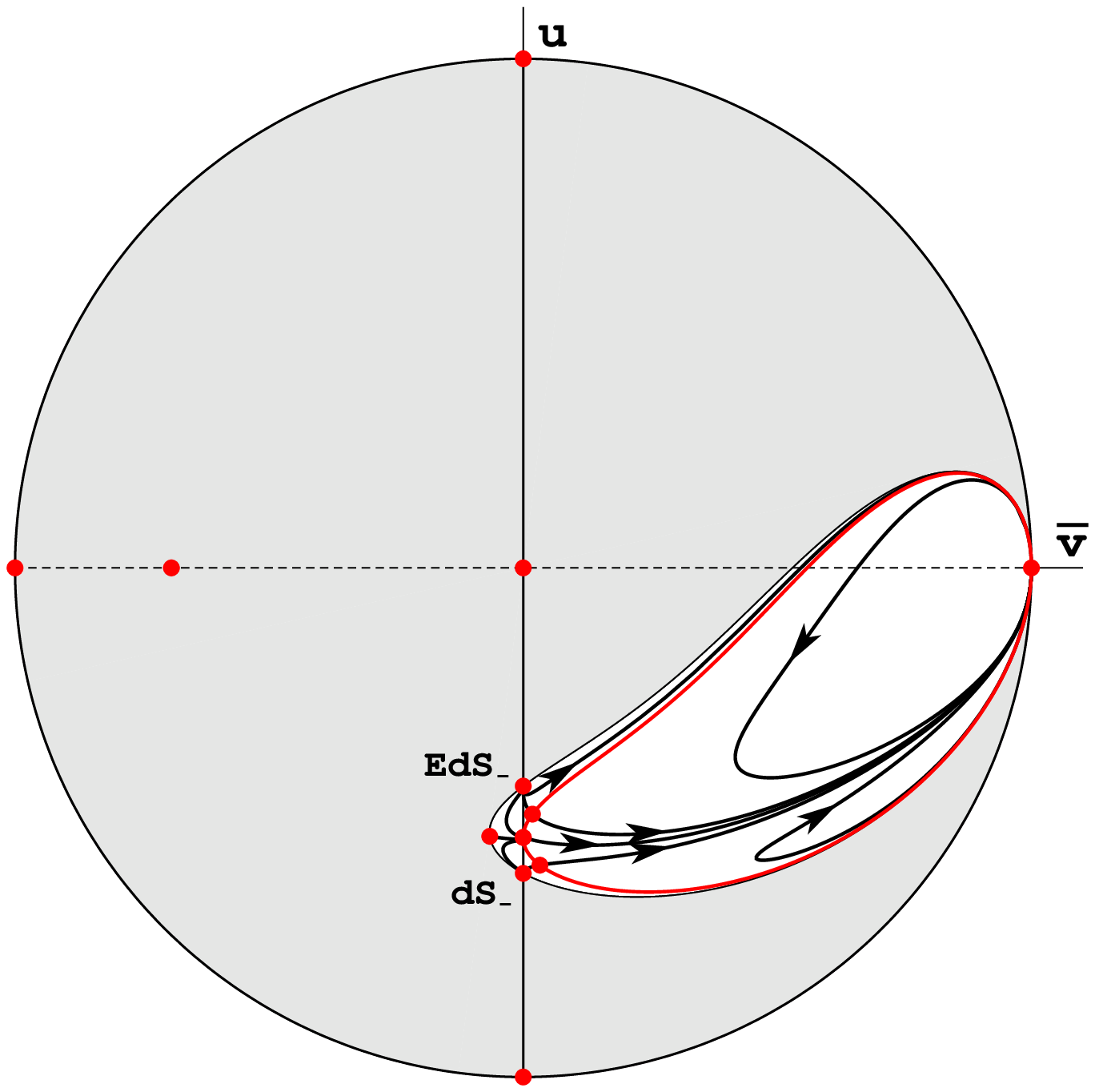}&
\includegraphics[scale=0.425]{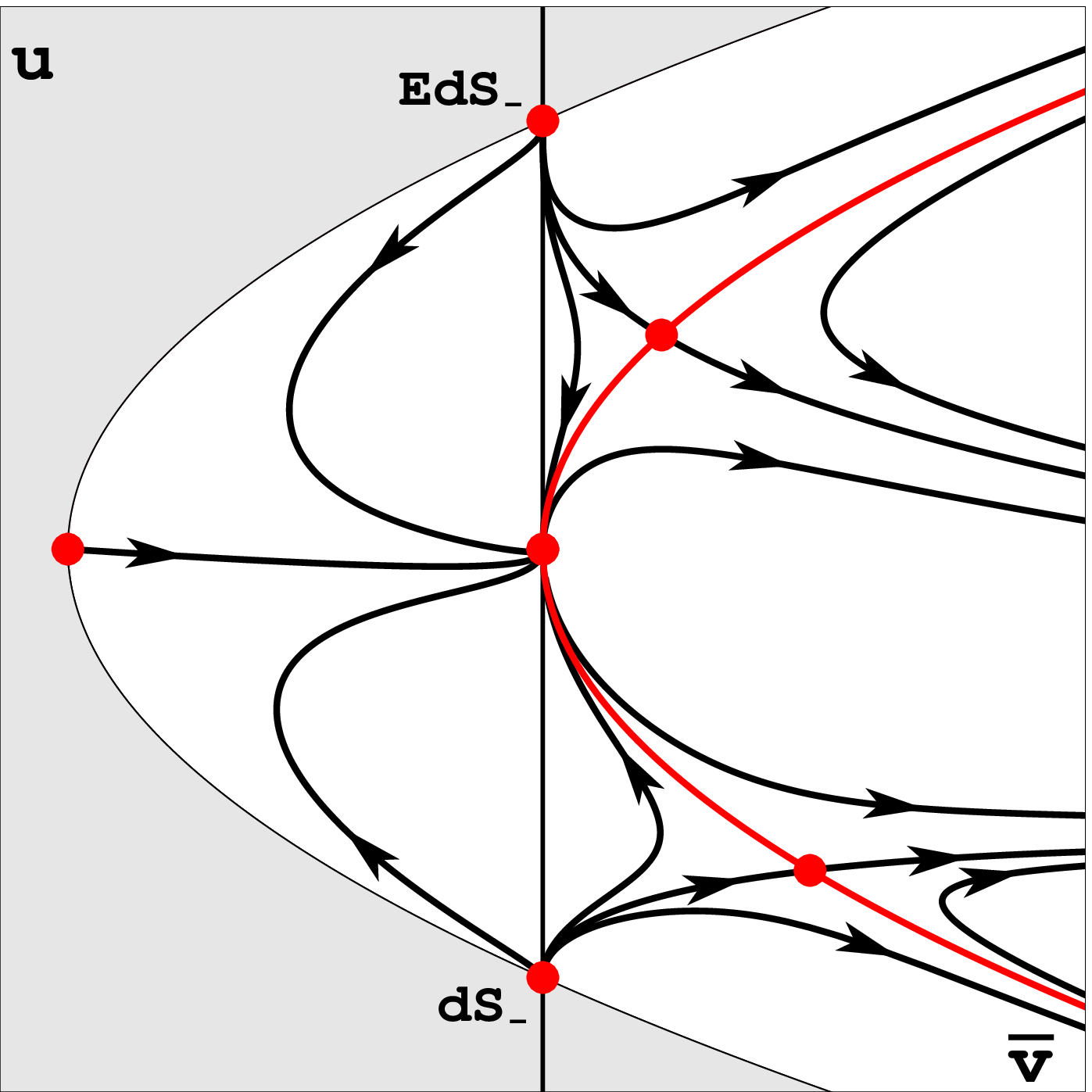}
\end{tabular}
\caption{The phase space diagrams representing evolutional paths of the decoupled part in the $(u,\bar{v})$ dynamical variables of the system \eqref{eq:dynsys_reduced} with $\lambda=2$ for the canonical $\ve=+1$ (top) and the phantom $\ve=-1$ (bottom) scalar fields. The shaded regions of the phase space are unphysical where the energy conservation condition \eqref{eq:hubble_arb_inf} is negative. Direction of arrows on the phase space trajectories indicates expansion of the universe. For $\lambda>-1$ both the Einstein-de Sitter state $EdS_{-}$ and the de Sitter state $dS_{-}$ are unstable with respect to expansion of the universe.}
\label{fig:1}
\end{figure*}

\begin{figure*}
\centering
\begin{tabular}{cc}
\includegraphics[scale=0.75]{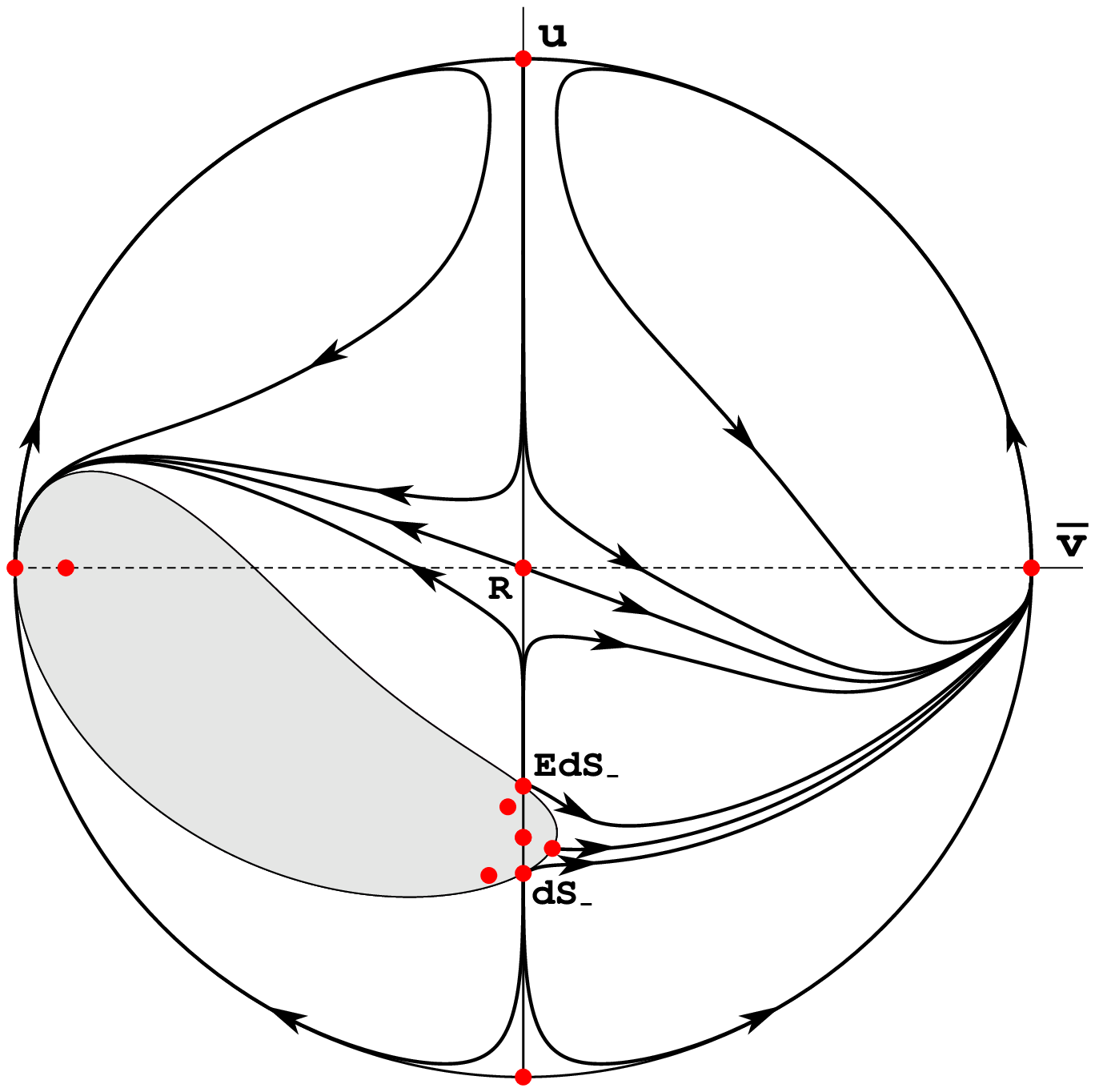} & \\
\includegraphics[scale=0.75]{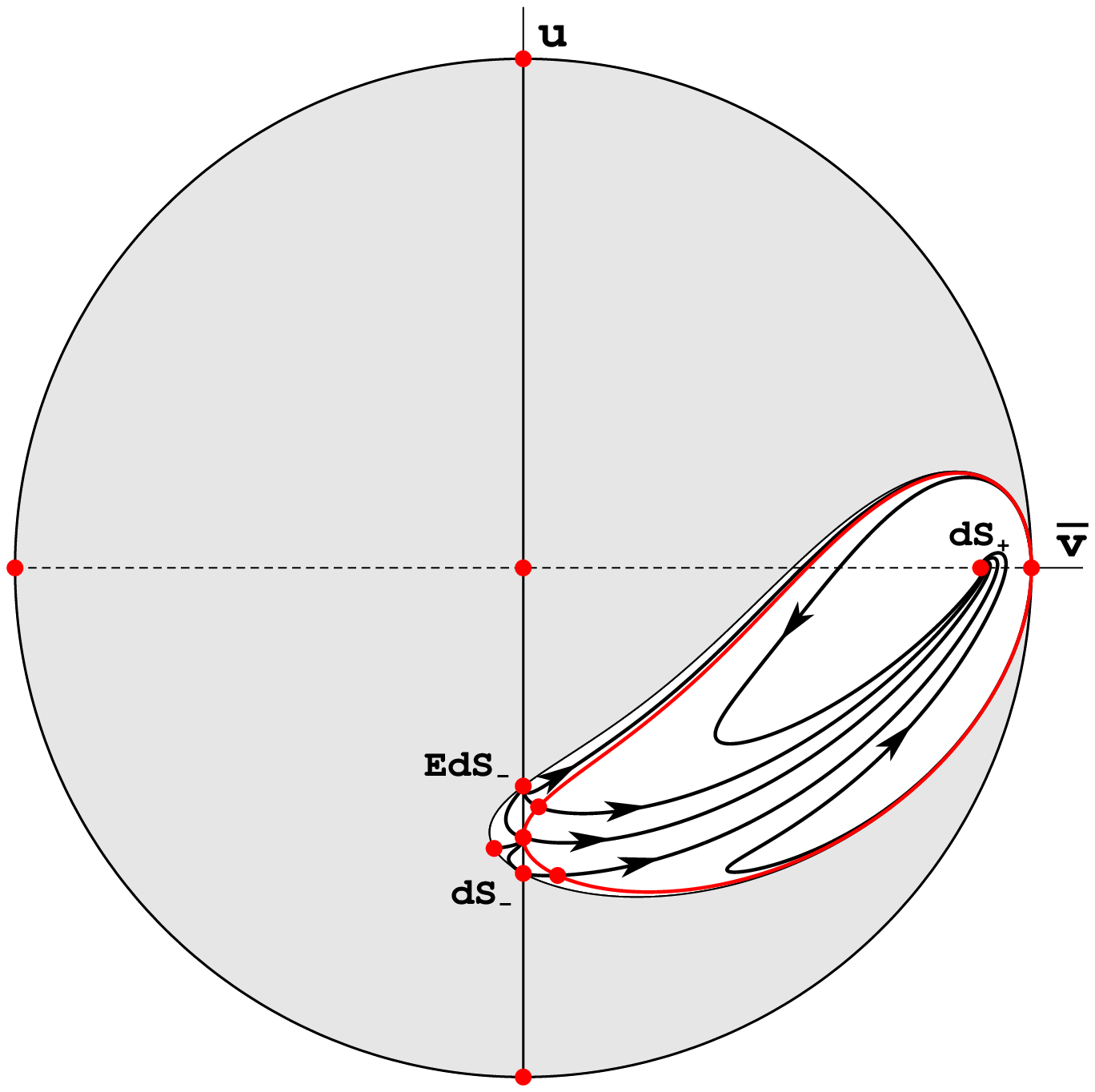}&
\includegraphics[scale=0.425]{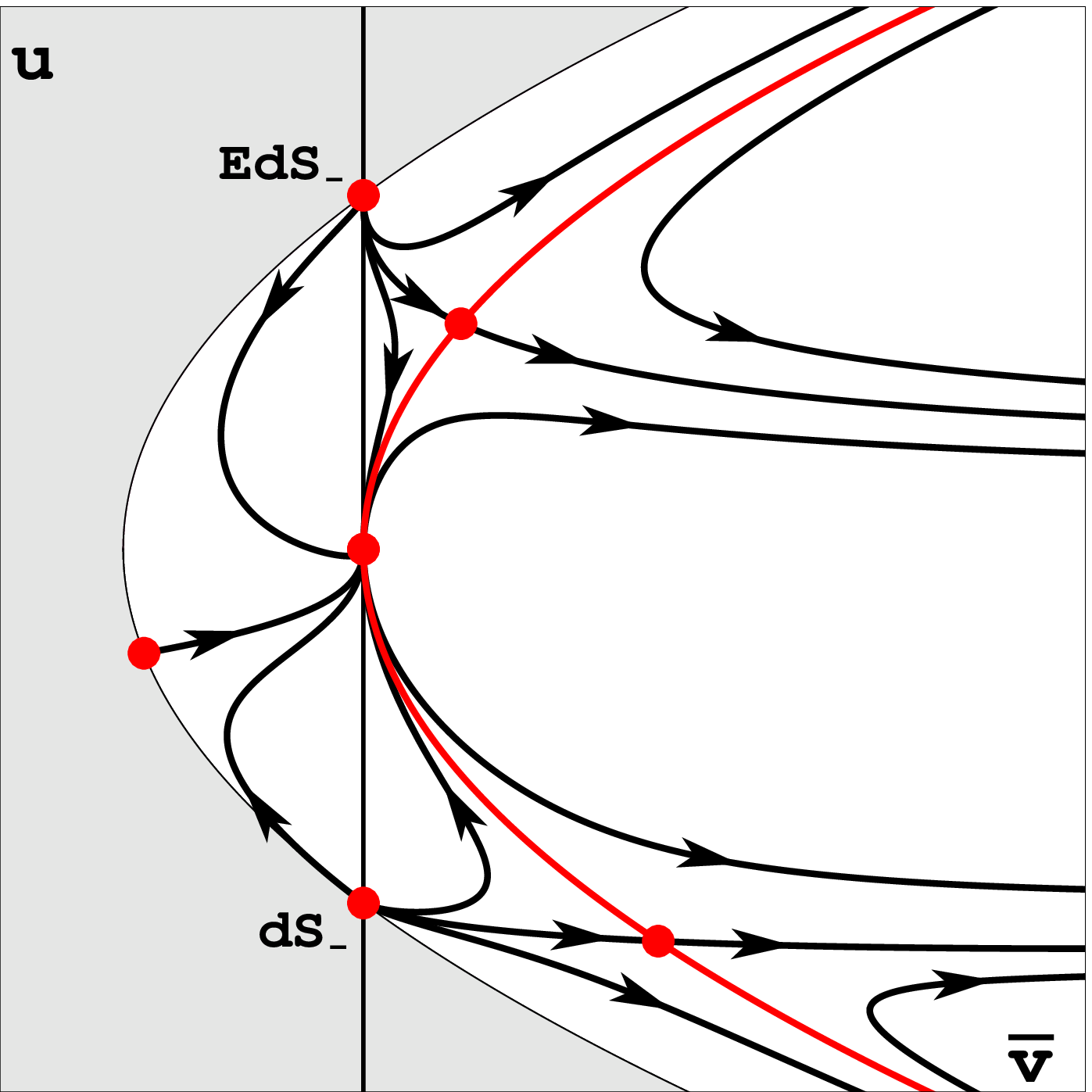}
\end{tabular}
\caption{The phase space diagrams representing evolutional paths of the decoupled part in the $(u,\bar{v})$ dynamical variables of the system \eqref{eq:dynsys_reduced} with $\lambda=-\frac{1}{2}$ for the canonical $\ve=+1$ (top) and the phantom $\ve=-1$ (bottom) scalar fields. The shaded regions of the phase space are unphysical where the energy conservation condition \eqref{eq:hubble_arb_inf} is negative. Direction of arrows on the phase space trajectories indicates expansion of the universe.  For $\lambda>-1$ both the Einstein-de Sitter state $EdS_{-}$ and the de Sitter state $dS_{-}$ are unstable with respect to expansion of the universe.}
\label{fig:2}
\end{figure*}

\begin{figure*}
\centering
\begin{tabular}{cc}
\includegraphics[scale=0.75]{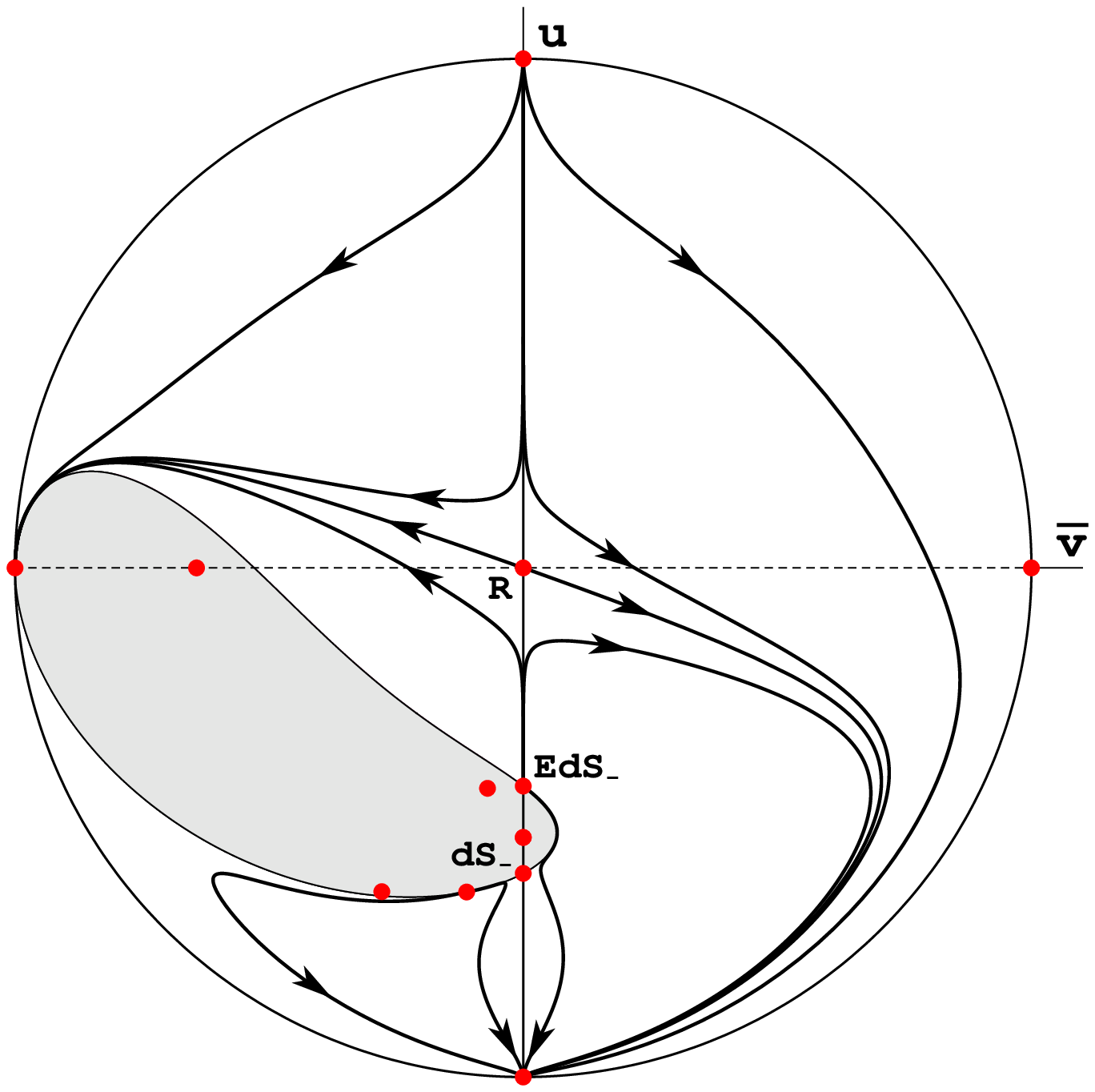} & \\
\includegraphics[scale=0.75]{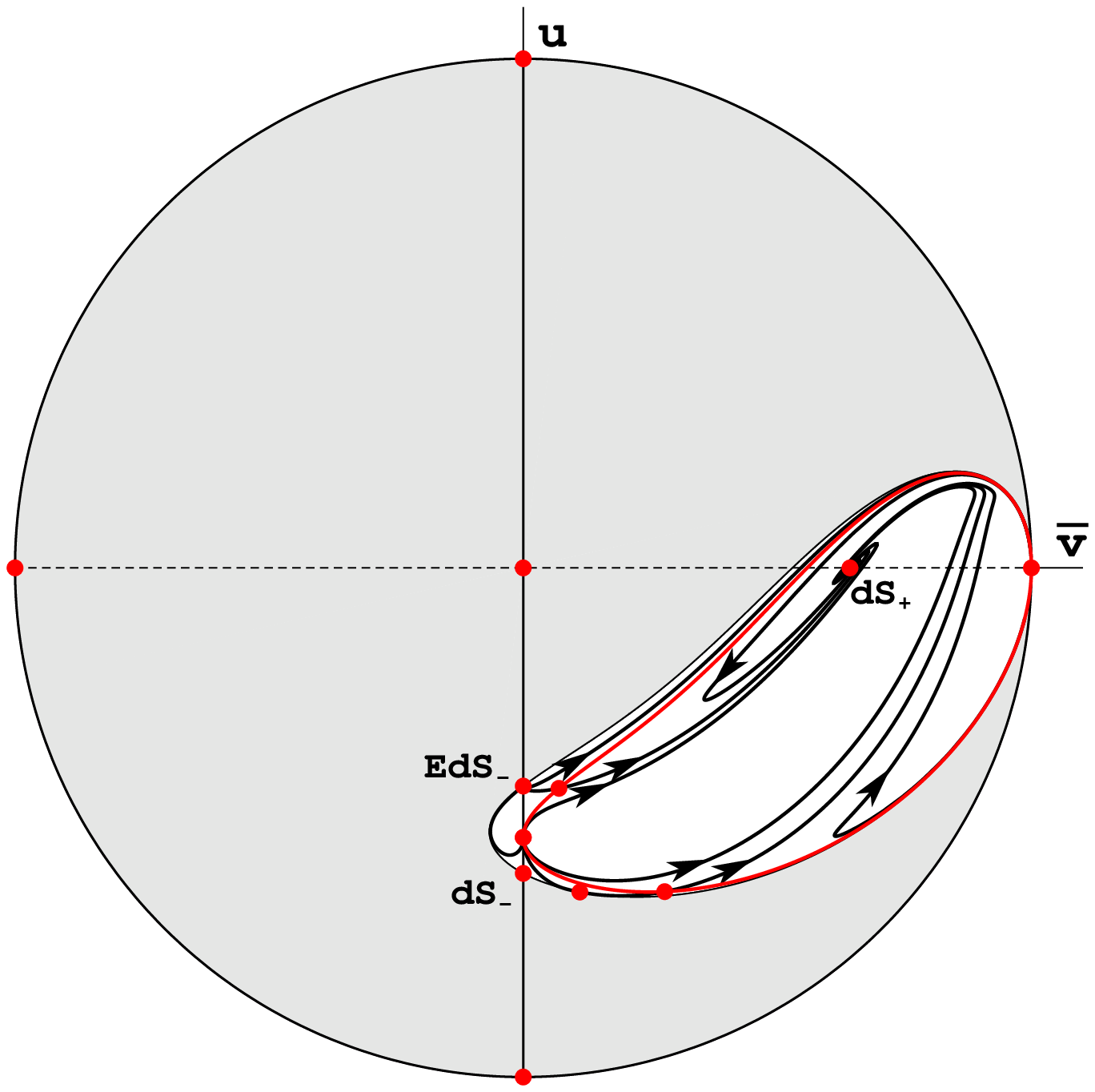}&
\includegraphics[scale=0.425]{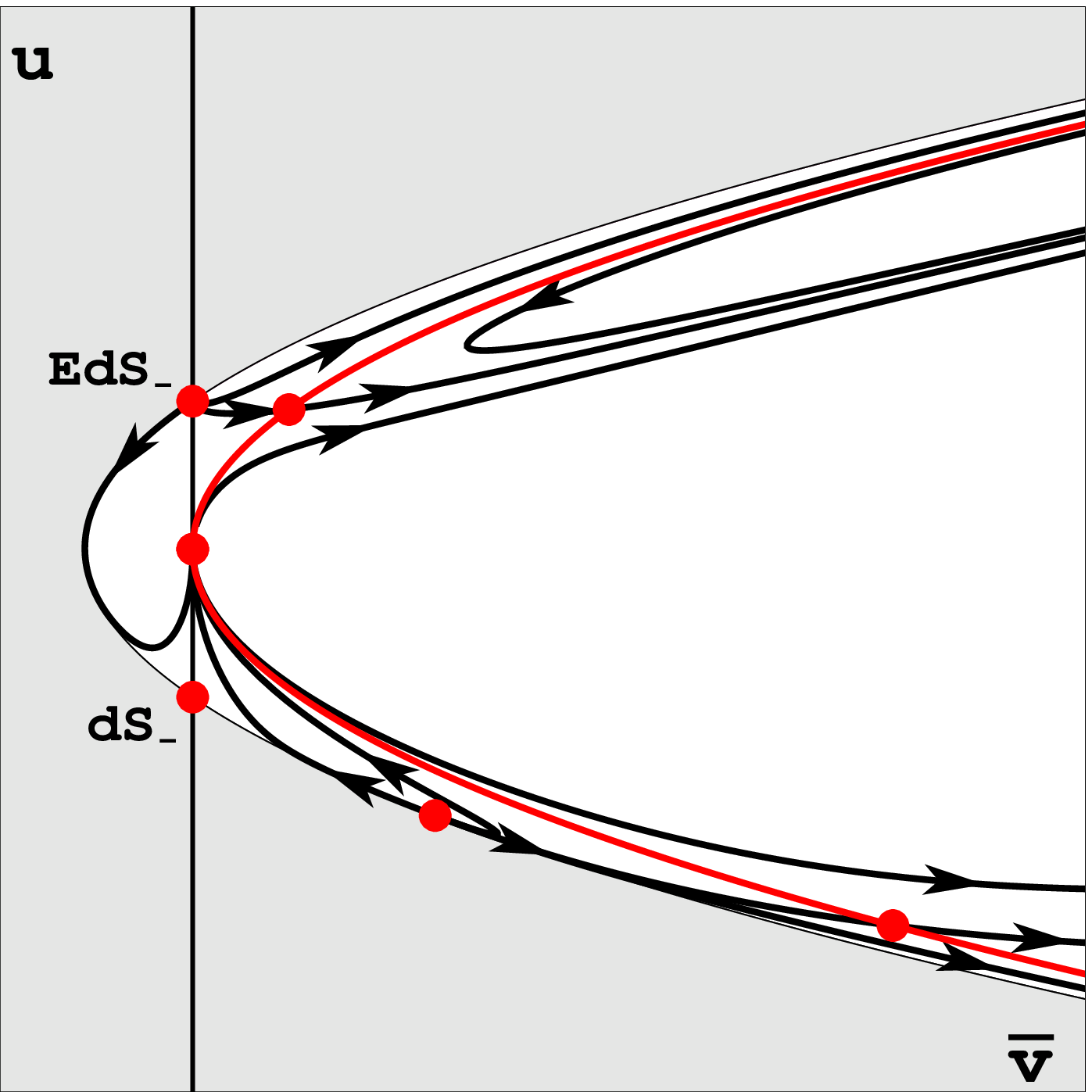}
\end{tabular}
\caption{The phase space diagrams representing evolutional paths of the decoupled part in the $(u,\bar{v})$ dynamical variables of the system \eqref{eq:dynsys_reduced} with $\lambda=-\frac{5}{2}$ for the canonical $\ve=+1$ (top) and the phantom $\ve=-1$ (bottom) scalar fields. The shaded regions of the phase space are unphysical where the energy conservation condition \eqref{eq:hubble_arb_inf} is negative. Direction of arrows on the phase space trajectories indicates expansion of the universe.  For $-5>\lambda>-1$ the Einstein-de Sitter state $EdS_{-}$ is unstable with respect to expansion of the universe while the de Sitter state $dS_{-}$ is in the form of a saddle type critical point.} 
\label{fig:3}
\end{figure*}

\begin{figure*}
\centering
\begin{tabular}{cc}
\includegraphics[scale=0.75]{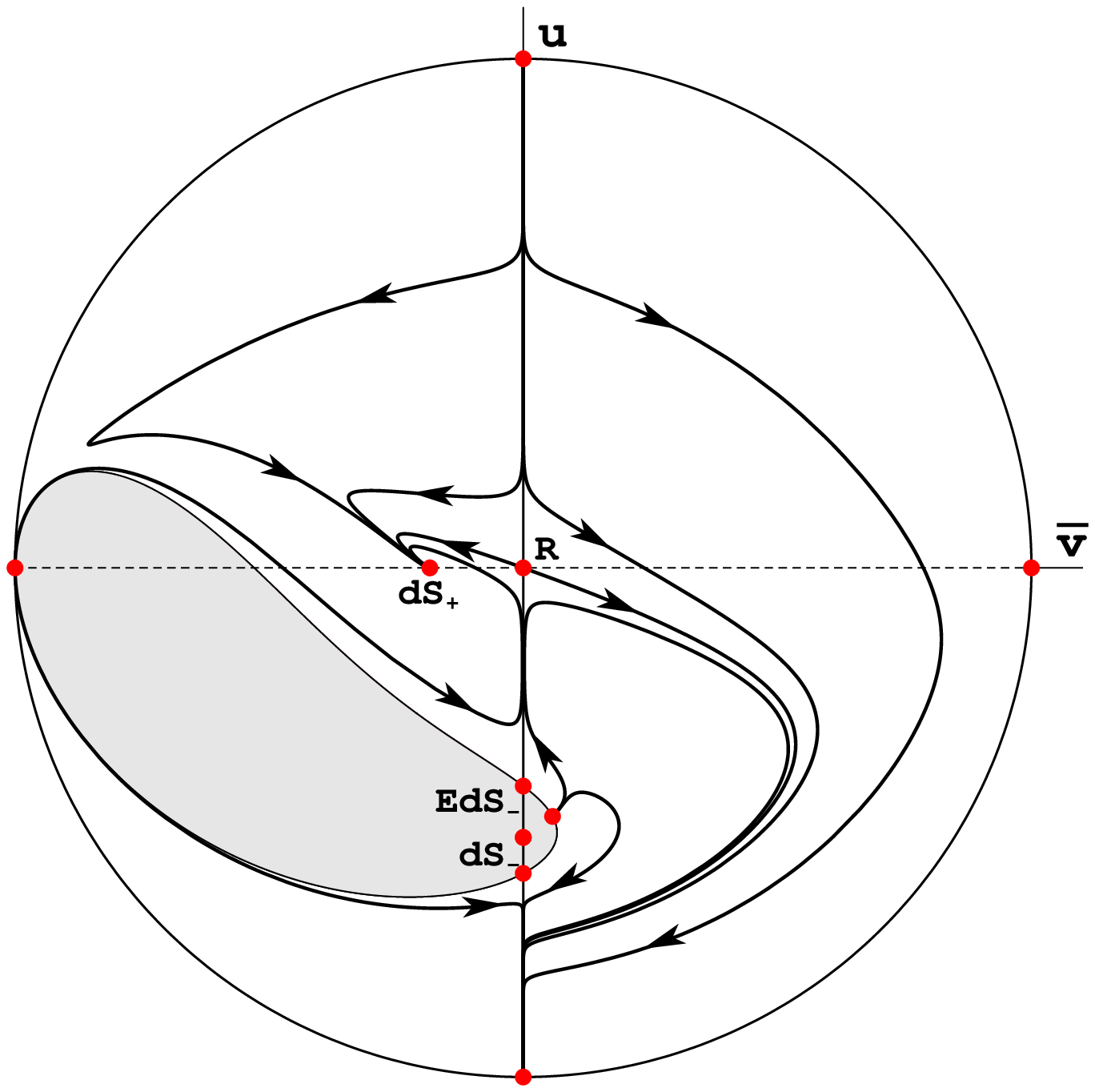} & \\
\includegraphics[scale=0.75]{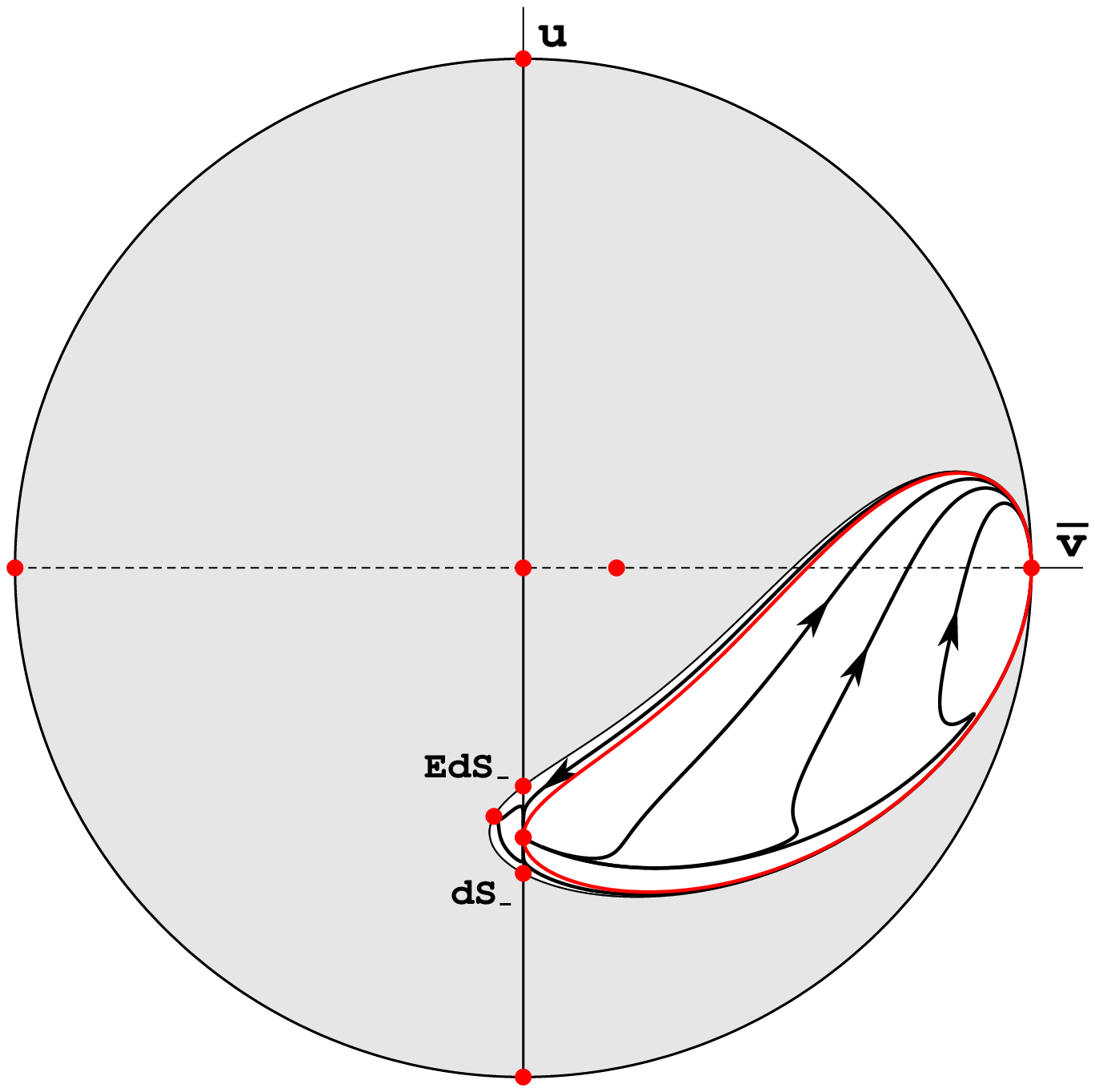}&
\includegraphics[scale=0.425]{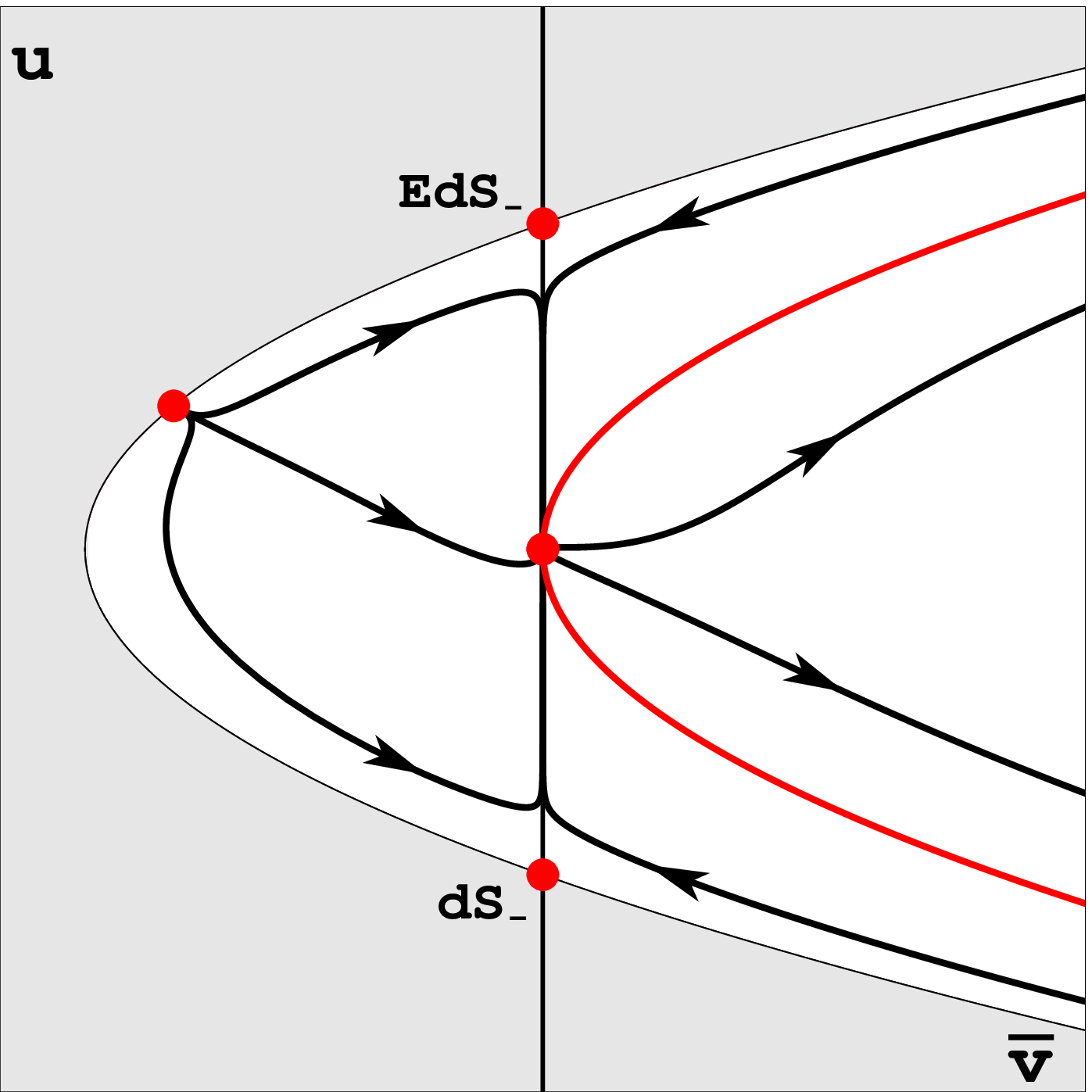}
\end{tabular}
\caption{The phase space diagrams representing evolutional paths of the decoupled part in the $(u,\bar{v})$ dynamical variables of the system \eqref{eq:dynsys_reduced} with $\lambda=-20$ for the canonical $\ve=+1$ (top) and the phantom $\ve=-1$ (bottom) scalar fields. The shaded regions of the phase space are unphysical where the energy conservation condition \eqref{eq:hubble_arb_inf} is negative. Direction of arrows on the phase space trajectories indicates expansion of the universe. For $\lambda<-5$ both states the Einstein-de Sitter $EdS_{-}$ and the de Sitter $dS_{-}$ are in the form of saddle type critical points. Note existence of the de Sitter state $dS_{+}$ for the canonical scalar field $\ve=+1$ which is asymptotically stable for negative potential functions.}
\label{fig:4}
\end{figure*}

In Figs.~ \ref{fig:1}, \ref{fig:2}, \ref{fig:3} and \ref{fig:4} we present four representative cases for dynamical behaviour for globally monomial scalar field potential functions in the $(u,\bar{v})$ variables compactified with the circle at infinity of the phase space. The shaded regions are unphysical and direction of the arrows indicates direction of expansion of the universe. In our discussion the most interesting is evolution in vicinity of the de Sitter and the Einstein-de Sitter states located at $\bar{v}^{*}=0$ because of the simplifying assumptions about global behaviour of the scalar field potential function which in most general case can not be true.  

The dynamical variable $\bar{v}$ is defined as $$\bar{v}=2\frac{U(\phi)}{H^{2}\phi^{2}}\,,$$ and on all the phase space diagrams the regions where $\bar{v}<0$ correspond to negative scalar field potential functions $U(\phi)<0$. It is interesting that the asymptotic unstable with respect to expansion of the universe de Sitter and Einstein-de Sitter states exist both for negative and positive potential functions. We have the following three generic cases : for $\lambda>-1$ both the Einstein-de Sitter and the de Sitter states are unstable with respect to the expansion of the universe (Figs.~\ref{fig:1} and \ref{fig:2}); for $-5>\lambda>-1$ the Einstein-de Sitters state corresponds to an unstable node while the de Sitter state is in the form of a saddle critical point (Fig.~\ref{fig:3}); for $\lambda<-5$ both state are in the form of saddle type critical points (Fig.~\ref{fig:4}). Additionally we have found that for a negative potential function there is a possibility for an asymptotically stable de Sitter state (see Fig.~\ref{fig:4}) \cite{Felder:2002jk,Boisseau:2015hqa}.

\section{A constant potential function}

In this section we present the full phase space analysis of the model with a constant potential function $U(\phi)=U_{0}=\text{const.}$ In such case the dynamical system describing the evolution of the phase space variables reduces to a $2$-dimensional dynamical system and is in the following form
\begin{equation}
\label{eq:const_fin}
\begin{split}
\frac{\ud x}{\ud\ln{a}} &= -3x-\frac{9}{8} z\bigg(\frac{\dot{H}}{H^{2}}+2\bigg)\,,\\
\frac{\ud z}{\ud\ln{a}} &= x+z\frac{\dot{H}}{H^{2}}\,,
\end{split}
\end{equation}
where the acceleration equation is given by
\begin{equation}
\label{eq:accel_fin}
\frac{\dot{H}}{H^{2}} = -2 + \frac{\ve\frac{1}{8}x^{2}+2\Omega_{\Lambda,0}}{\Omega_{\Lambda,0}+\ve\left(x+\frac{9}{8} z\right)^{2}}\,,
\end{equation}
the energy conservation condition is
\begin{equation}
\label{eq:hubble_fin}
h^{2} = \left(\frac{H}{H_{0}}\right)^{2} = \Omega_{\Lambda,0} + \ve\left(x+\frac{3}{4}z\right)\left(x+\frac{3}{2}z\right)\ge0\,,
\end{equation}
and the constant $\Omega_{\Lambda,0}=\frac{\kappa^{2}U_{0}}{3H_{0}^{2}}$.

The last inequality defines the physical regions of the phase space variables. In general case with the non-minimal coupling constant as a free parameter shape of this region crucially depends on the value of the non-minimal coupling constant $\xi$ as well as on the type of the scalar field.

The right hand sides of the system \eqref{eq:const_fin} are rational functions of the phase space variables due to the form of the acceleration equation \eqref{eq:accel_fin}. Such dynamical system provides us only with partial information about asymptotic states of dynamics, we are unable to obtain asymptotic states for which denominator in the acceleration equation \eqref{eq:accel_fin} vanishes. Using the following time transformation
\begin{equation}
\label{eq:time_rep}
\frac{\ud}{\ud\tau} = \left(\Omega_{\Lambda,0}+\ve\left(x+\frac{9}{8}z\right)^{2}\right)\frac{\ud}{\ud\ln{a}}\,,
\end{equation}
we obtain the dynamical system \eqref{eq:const_fin} in the form
\begin{equation}
\label{eq:const_fin_2}
\begin{split}
\frac{\ud x}{\ud\tau} & = -3\Omega_{\Lambda,0}\left(x+\frac{3}{4} z\right) \\ & \hspace{4.5mm} -
\ve3x\left(\left(x+\frac{9}{8} z\right)^{2}+\frac{3}{64}x z\right)\,,\\
\frac{\ud z}{\ud\tau} & =\Omega_{\Lambda,0}x+\ve\left((x-2z)\left(x+\frac{9}{8} z\right)^{2}+ \frac{1}{8}x^{2}z\right)\,,
\end{split}
\end{equation}
where now the right hand sides are polynomials in the phase space variables.

\begin{table*}
\caption{The critical points of the system \eqref{eq:const_fin_2} and corresponding values of the Hubble function \eqref{eq:hubble_fin} and the acceleration equation \eqref{eq:accel_fin}.}
\label{tab:2}
\renewcommand{\arraystretch}{1.5}
	\centering
	\begin{tabular}{|c|l|l|c|c|c|}
		\hline
		& $x^{*}$ & $z^{*}$ & $\big(\frac{H}{H_{0}}\big)^{2}\big|^{*}$ & $\frac{\dot{H}}{H^{2}}\big|^{*}$ & $\frac{\dot{H}}{H^{2}_{0}}\big|^{*}$\\
		\hline
		$\text{dS}_{+}$ & $0$ & $0$ & $\Omega_{\Lambda,0}$ & $0$  & $0$ \\
		E & $\pm4\sqrt{\ve\frac{2}{3}\Omega_{\Lambda,0}}$ & $\mp\frac{10}{3}\sqrt{\ve\frac{2}{3}\Omega_{\Lambda,0}}$ & $0$ & $\frac{6}{5}$ & $0$ \\
		B, K & $\pm\sqrt{-\ve\Omega_{\Lambda,0}}$ & $0$ & $0$ & $\pm\infty$ & $\frac{15}{8}\Omega_{\Lambda,0}$ \\
		d & $\pm4\sqrt{-\ve\Omega_{\Lambda,0}}$ &
		$\mp\frac{8}{3}\sqrt{-\ve\Omega_{\Lambda,0}}$ & $\Omega_{\Lambda,0}$ & $0$ & $0$\\
		e & $\pm4\sqrt{-\ve\Omega_{\Lambda,0}}$ & $\mp\frac{40}{9}\sqrt{-\ve\Omega_{\Lambda,0}}$ & $\frac{25}{9}\Omega_{\Lambda,0}$ & $-\frac{4}{5}$& $-\frac{20}{9}\Omega_{\Lambda,0}$\\
		\hline
	\end{tabular}
\end{table*}

The critical points of the system \eqref{eq:const_fin_2} are presented in Table \ref{tab:2} together with the corresponding values of the Hubble function \eqref{eq:hubble_fin} and the acceleration equation \eqref{eq:accel_fin} calculated at those asymptotic states. Note that the critical points $dS_{+}$ and $E$ are the asymptotic states for both systems \eqref{eq:const_fin} and \eqref{eq:const_fin_2} while the remaining points only for the system \eqref{eq:const_fin_2}. This is the reason why, in what follows, we present stability analysis of the critical points $dS_{+}$ and $E$ analysing the system \eqref{eq:const_fin} where the ``time'' parameter along the phase space curves is the scale factor. It will provide us with natural interpretation of the stability conditions with respect to the expansion of universe.

\begin{figure*}
\centering
\includegraphics[scale=0.75]{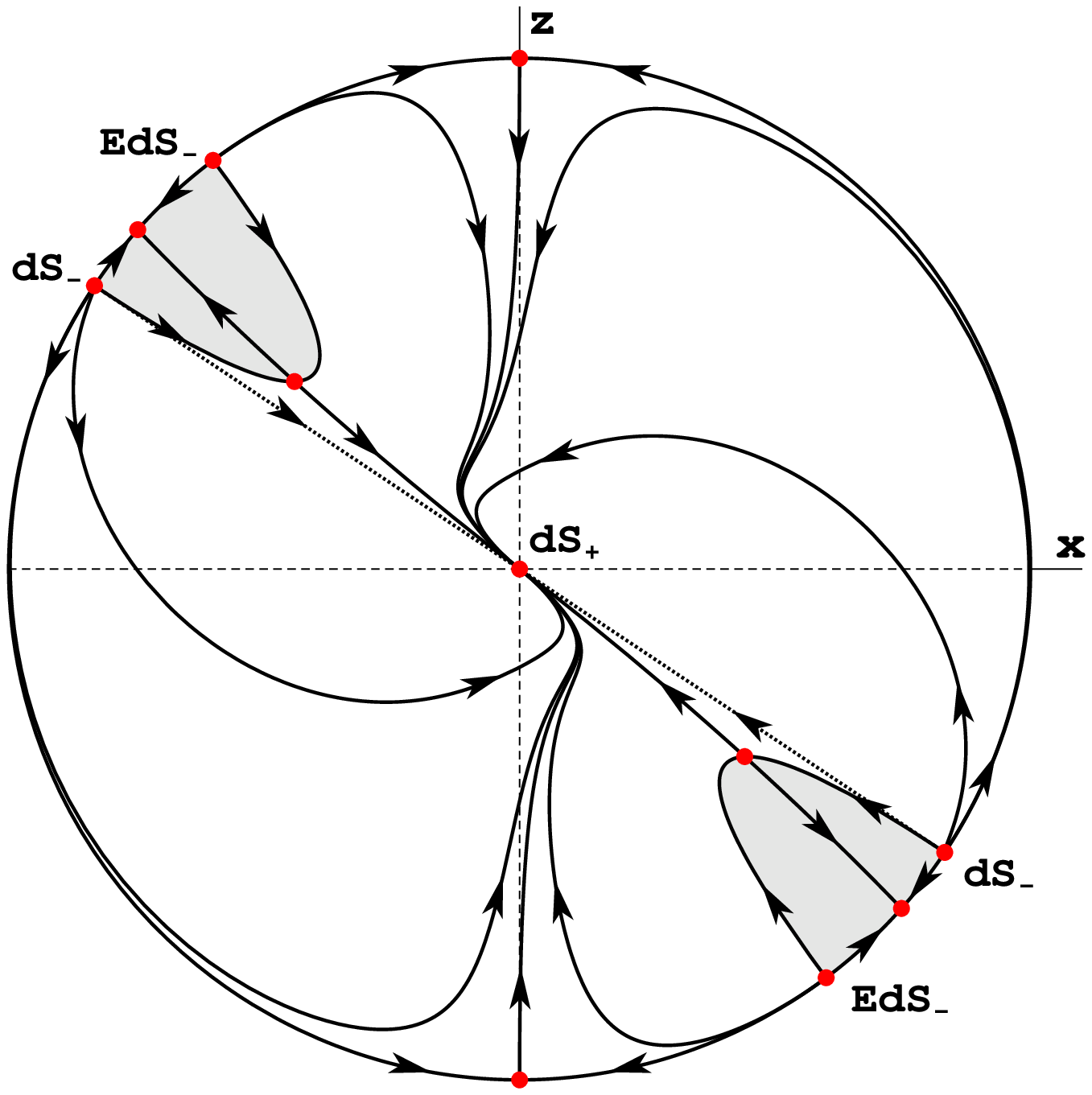}
\includegraphics[scale=0.75]{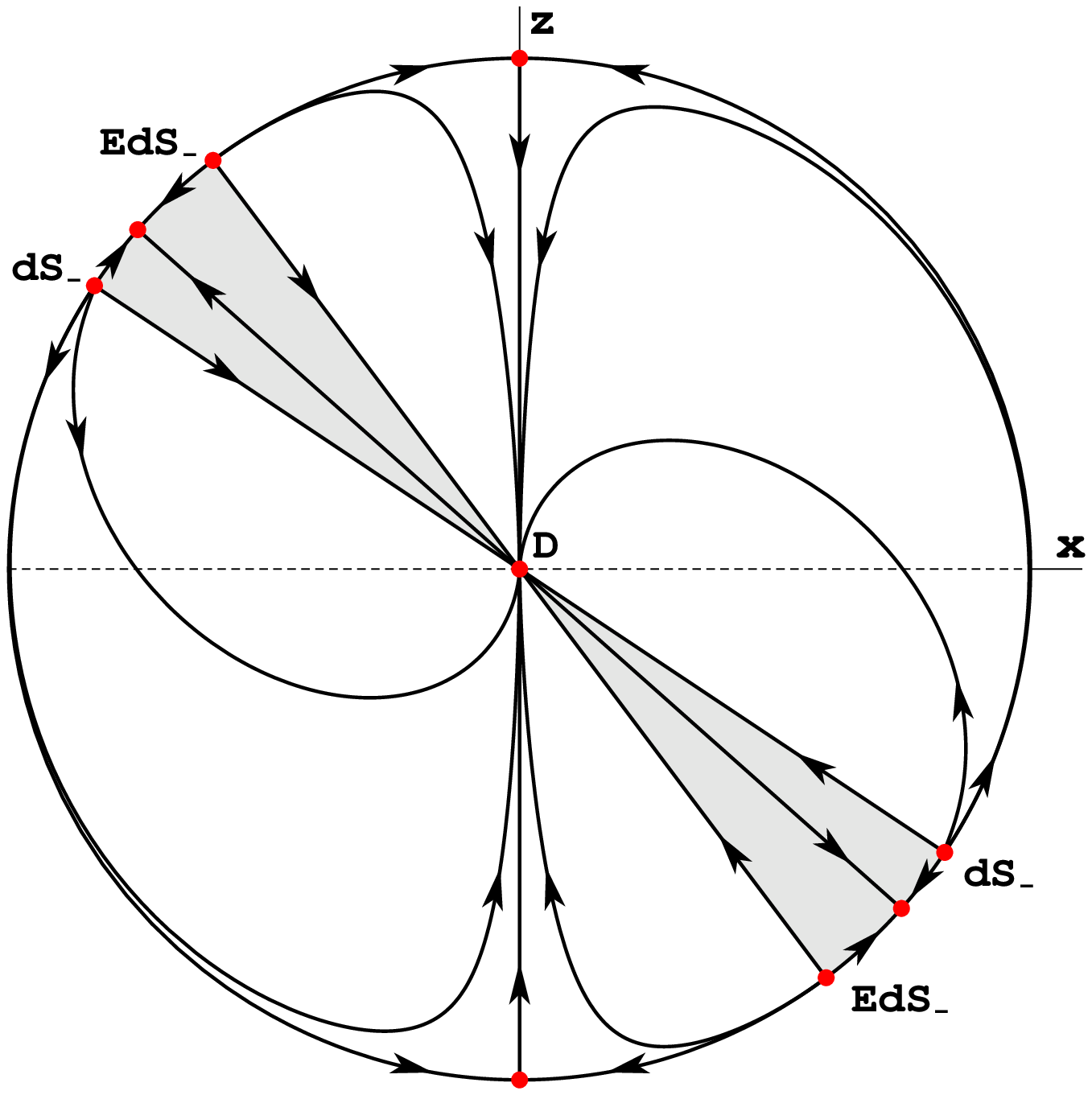}
\caption{The phase space diagrams representing evolutional paths in the phase space compactified with the circle at infinity for models with a canonical scalar field $\ve=+1$ and $\xi=\frac{3}{16}$, $\Omega_{\Lambda,0}>0$ (top); and with $\Omega_{\Lambda,0}=0$ (bottom). The shaded regions where the energy conservation condition \eqref{eq:hubble_fin} is negative are nonphysical. The direction of arrows in the physical regions of the phase space corresponds to expansion of universe. The dotted trajectory on the first diagram corresponds to the pure de Sitter type of evolution. On the bottom diagram at the centre we can observe that the critical points from the previous diagram merge giving rise to a degenerated critical points with an asymptotically stable regions in the physical parts of the phase space. The vanishing cosmological constant $\Omega_{\Lambda,0}=0$ is the bifurcation value for dynamics of the model.}
\label{fig:5}
\end{figure*}

\begin{figure*}
\centering
\includegraphics[scale=0.75]{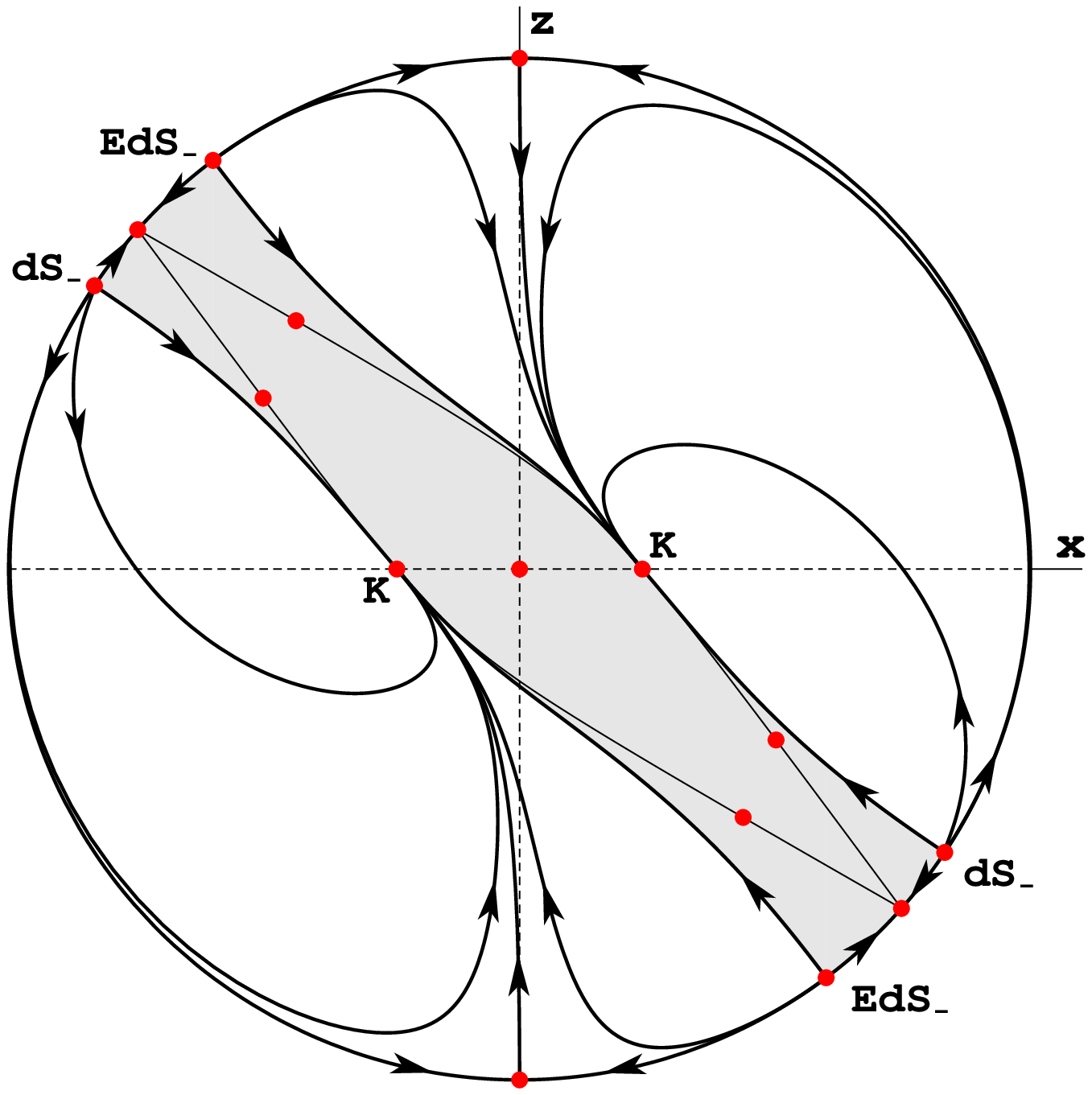}
\includegraphics[scale=0.75]{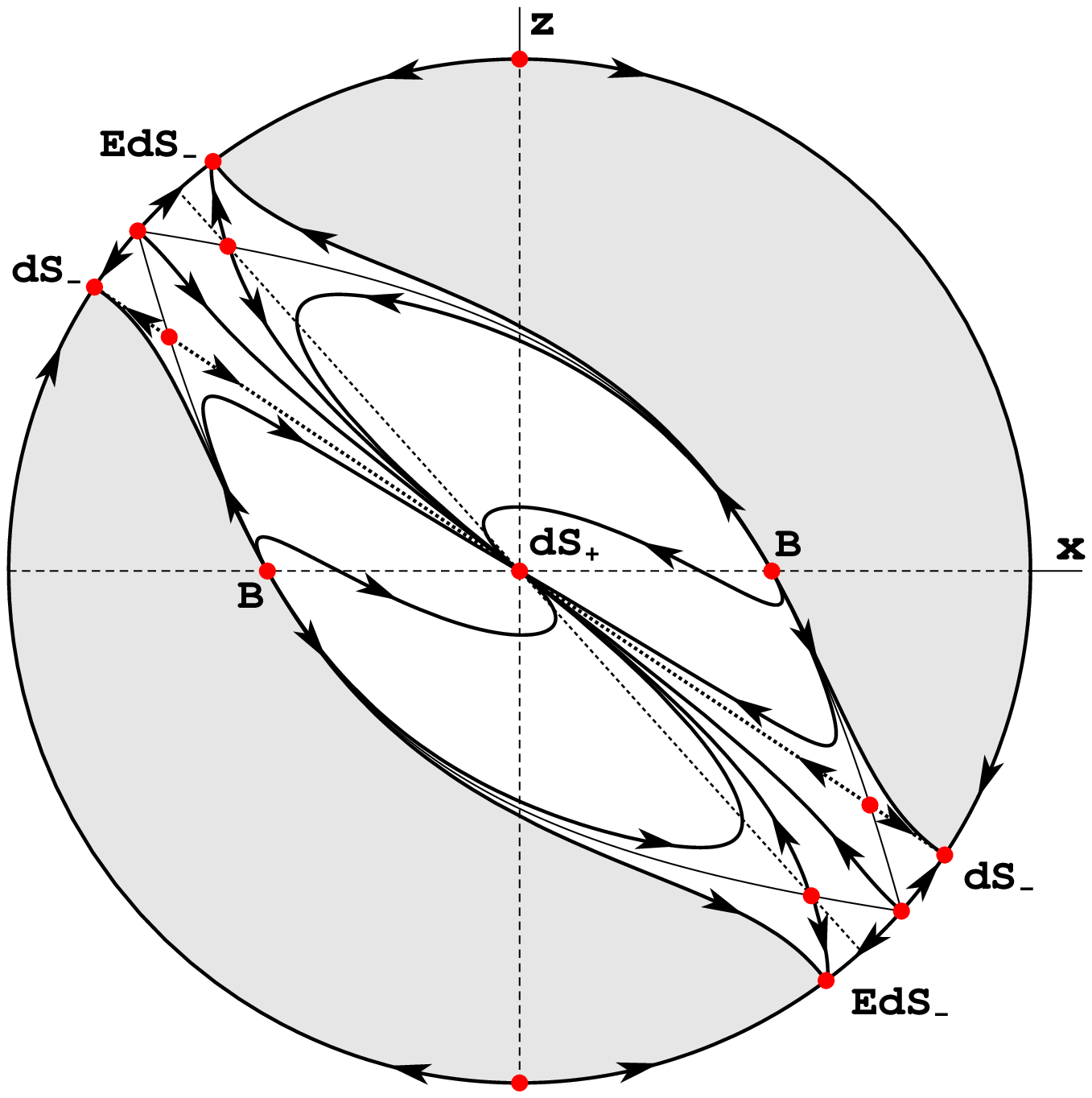}
\caption{The phase space diagrams representing evolutional paths in the phase space compactified with the circle at infinity for models with the canonical scalar field $\ve=+1$ and $\xi=\frac{3}{16}$, $\Omega_{\Lambda,0}<0$ (top); and the phantom scalar field $\ve=-1$ and $\xi=\frac{3}{16}$, $\Omega_{\Lambda,0}>0$ (bottom). The shaded regions where the energy conservation condition \eqref{eq:hubble_fin} is negative are nonphysical. Note a complementarity between evolutional paths in a physical regions of the phase portraits. }
\label{fig:6}
\end{figure*}

The first critical point $$(x^{*}=0\,,z^{*}=0)$$ corresponds to the de Sitter type of evolution and is denoted as $dS_{+}$ in Table \ref{tab:2} and on phase space portraits \ref{fig:5} and \ref{fig:6}. The acceleration equation \eqref{eq:accel_fin} calculated at this asymptotic state vanishes and the energy conservation condition \eqref{eq:hubble_fin} gives value of the Hubble function at this state
\begin{equation}
\left(\frac{H}{H_{0}}\right)^{2}\Bigg|^{*} = \left(h^{*}\right)^{2} = \Omega_{\Lambda,0}\,,
\end{equation}
and the following condition must be fulfilled $\Omega_{\Lambda,0}>0$ in order to obtain the asymptotic state in the physical region of the phase space. The stability conditions may be obtained directly from the eigenvalues of the linearisation matrix of the system. They are
\begin{equation}
\lambda_{1}=\lambda_{2}=-\frac{3}{2}\,,
\end{equation}
which indicates that the asymptotic de Sitter state exists in form of a stable node. The linearised solutions to dynamical system \eqref{eq:const_fin} in the vicinity of this state are
\begin{equation}
\begin{split}
x(a) & = \Delta x \left(\frac{a}{a^{(i)}}\right)^{-\frac{3}{2}} \\ & \hspace{4mm} - \frac{3}{2}\left(\Delta x+\frac{3}{2}\Delta z\right)\left(\frac{a}{a^{(i)}}\right)^{-\frac{3}{2}}\ln{\left(\frac{a}{a^{(i)}}\right)}\,,\\
z(a) & = \Delta z \left(\frac{a}{a^{(i)}}\right)^{-\frac{3}{2}} \\ & \hspace{4mm} + \left(\Delta x+\frac{3}{2}\Delta z\right)\left(\frac{a}{a^{(i)}}\right)^{-\frac{3}{2}}\ln{\left(\frac{a}{a^{(i)}}\right)}\,,
\end{split}
\end{equation} 
where $\Delta x = x^{(i)}$, $\Delta z = z^{(i)}$ are initial conditions for the phase space variables and $a^{(i)}$ is initial value of the scale factor. Then, using those linearised solutions we can obtain the Hubble function from \eqref{eq:hubble_fin} up to second order terms in initial conditions
\begin{equation}
\begin{split}
\left(\frac{H(a)}{H(a_{0})}\right)^{2} & = \Omega_{\Lambda,0} \\ & + \ve\left(\Delta x +\frac{3}{2} \Delta z\right)\left(\Delta x + \frac{3}{4}\Delta z\right)\left(\frac{a}{a^{(i)}}\right)^{-3} \\ & -
\ve\frac{3}{4}\left(\Delta x + \frac{3}{2}\Delta z\right)^{2}\left(\frac{a}{a^{(i)}}\right)^{-3}
\ln{\left(\frac{a}{a^{(i)}}\right)}\,,
\end{split}
\end{equation}
which resembles the Hubble function for the $\Lambda$CDM model except the term logarithmic in the scale factor. It is easy to show that the system \eqref{eq:const_fin} is equipped with an 
invariant manifold of the de Sitter type of evolution $x+\frac{3}{2}z\equiv0$ where the acceleration equation \eqref{eq:accel_fin} vanishes and the Hubble function is constant. Then we can assume that we are in close vicinity of this state $\left(\Delta x +\frac{3}{2} \Delta z\right)^{2}\approx0$ then the Hubble function is
\begin{equation}
\left(\frac{H(a)}{H(a_{0})}\right)^{2} \approx \Omega_{\Lambda,0} + \ve\left(\Delta x +\frac{3}{2} \Delta z\right)\left(\Delta x + \frac{3}{4}\Delta z\right)\left(\frac{a}{a^{(i)}}\right)^{-3}\,,
\end{equation}
which effectively corresponds to $\Lambda$CDM model. 

The invariant manifold trajectories corresponding to the de Sitter type of evolution are presented as a dotted trajectories at the top panel Fig.~\ref{fig:5} and at the bottom panel Fig.~\ref{fig:6}.

The next two critical points denoted as $E$ in Table \ref{tab:2} are located at $$\left(x^{*}=\pm4\sqrt{\ve\frac{2}{3}\Omega_{\Lambda,0}}\,, z^{*}=\mp\frac{10}{3}\sqrt{\ve\frac{2}{3}\Omega_{\Lambda,0}}\right)$$ with vanishing energy conservation condition \eqref{eq:hubble_fin} and the acceleration equation \eqref{eq:accel_fin} 
$\frac{\dot{H}}{H^{2}}\big|^{*}=\frac{6}{5}$ give rise to the Einstein static universe, because first cosmological time derivative of the Hubble function calculated at this point vanishes $\frac{\dot{H}}{H_{0}^{2}}\big|^{*}=0$. We have to note that this state exists only when the following condition is fulfilled
\begin{equation}
\ve\Omega_{\Lambda,0}>0\,,
\end{equation}
they exist only for the canonical scalar field $\ve=+1$ with positive cosmological constant $\Omega_{\Lambda,0}>0$ and for the phantom scalar field $\ve=-1$ with the negative cosmological constant $\Omega_{\Lambda,0}<0$. 

The eigenvalues of the linearisation matrix of the system \eqref{eq:const_fin} calculated at those points are
\begin{equation}
\lambda_{1}=\frac{12}{5}\,,\quad\lambda_{2}=-\frac{9}{5}\,,
\end{equation}
which indicates that the Einstein static solution is in the form of a saddle type critical point. 

The linearised solutions to the dynamics in the vicinity of this state are the following
\begin{equation}
\begin{split}
x(a) & = x^{*} - \frac{22}{35}\left(\Delta x + 3 \Delta z\right)\left(\frac{a}{a^{(i)}}\right)^{\frac{12}{5}} \\ & + \frac{3}{35}\left(19\Delta x+22\Delta z\right)\left(\frac{a}{a^{(i)}}\right)^{-\frac{9}{5}}\,,\\
z(a) & = z^{*} + \frac{19}{35}\left(\Delta x + 3 \Delta z\right)\left(\frac{a}{a^{(i)}}\right)^{\frac{12}{5}} \\ & - \frac{1}{35}\left(19\Delta x+22\Delta z\right)\left(\frac{a}{a^{(i)}}\right)^{-\frac{9}{5}} \,,
\end{split}
\end{equation}
where $\Delta x = x^{(i)}-x^{*}$, $\Delta z = z^{(i)}-z^{*}$ are initial conditions for the phase space variables and $a^{(i)}$ is initial value of the scale factor. Using these solutions we obtain the Hubble function 
\begin{equation}
\left(\frac{H(a)}{H(a_{0})}\right)^{2} \approx \ve\sqrt{\ve\frac{1}{6}\Omega_{\Lambda,0}}\left(\Delta x + 3\Delta z\right)\left(\frac{a}{a^{(i)}}\right)^{\frac{12}{5}}\,,
\end{equation}
where the dependence on the scale factor is connected only with an unstable direction of the saddle type critical point and condition for evolution in the physical region of the phase space is $\ve\left(\Delta x + 3\Delta z\right)>0$.

Now we can proceed to investigate rest of the asymptotic states of the system exploring dynamical behaviour of the system \eqref{eq:const_fin_2} in reparametrised time variable. The first critical points, in Table \ref{tab:2} denoted as $B,K$ (depending on the physical type) have the following coordinates in the phase space
$$\left(x^{*}=\pm\sqrt{-\ve\Omega_{\Lambda,0}}\,, z^{*}=0\right)$$
with vanishing energy conservation condition \eqref{eq:hubble_fin} exists only when 
$\ve\Omega_{\Lambda,0}<0$, i.e., for the canonical scalar field $\ve=+1$ we have $\Omega_{\Lambda,0}<0$ and for the phantom scalar field $\ve=-1$ we have $\Omega_{\Lambda,0}>0$. The eigenvalues of the linearisation matrix calculated at this point are the following
\begin{equation}
\lambda_{1}=\frac{15}{4}\Omega_{\Lambda,0}\,,\quad \lambda_{2}=\frac{15}{8}\Omega_{\Lambda,0}\,,
\end{equation}
which are of the same sign and stability conditions in time $\tau$ depends on sign of the cosmological constant $\Omega_{\Lambda,0}$.

The linearised solution are
\begin{equation}
\begin{split}
x(\tau) & = x^{*} +\frac{11}{5}\left(\Delta x+\frac{9}{8}\Delta z\right)\exp{(\lambda_{1}\tau)} \\ &-\frac{3}{40}\left(16\Delta x +33\Delta z\right)\exp{(\lambda_{2}\tau)}\,,\\
z(\tau) & = -\frac{16}{15}\left(\Delta x+\frac{9}{8}\Delta z\right)\exp{(\lambda_{1}\tau)} \\ &+ \frac{1}{15}\left(16\Delta x +33\Delta z\right)\exp{(\lambda_{2}\tau)}\,,
\end{split}
\end{equation}
where $\Delta x = x^{(i)}-x^{*}$, $\Delta z = z^{(i)}$ are initial conditions for the phase space variables.

Up to linear terms in initial conditions the dynamical time reparameterisation \eqref{eq:time_rep} gives us differential equation for the scale factor
\begin{equation}
\ud\ln{a}\approx\ve2 x^{*}\left(\Delta x+\frac{9}{8}\Delta z\right)\exp{\left(\frac{15}{4}\Omega_{\Lambda,0}\tau\right)}\ud\tau\,,
\end{equation}
which directly connects the dynamical time and change of the cosmological scale factor.
After integration this equation and the Hubble function \eqref{eq:hubble_fin} constitute parametric solution to dynamics in the vicinity of the critical point
\begin{equation}
\left\{
\begin{split}
\left(\frac{H(a)}{H(a_{0})}\right)^{2} & \approx \ve2x^{*}\left(\Delta x+\frac{9}{8}\Delta z\right)\exp{\left(\frac{15}{4}\Omega_{\Lambda,0}\tau\right)}\,,\\
\ln{\left(\frac{a}{a^{(i)}}\right)} & \approx \ve\frac{8}{15\Omega_{\Lambda,0}}x^{*}\left(\Delta x +\frac{9}{8}\Delta z\right)\left(\exp{\left(\frac{15}{4}\Omega_{\Lambda,0}\tau\right)}-1\right)\,.
\end{split}
\right.
\end{equation}
The condition for evolution in the physical region of the phase space can be obtain from the positivity of the Hubble function as
\begin{equation}
\ve x^{*}\left(\Delta x+\frac{9}{8}\Delta z\right) >0\,.
\end{equation}

From \eqref{eq:accel_fin} and \eqref{eq:hubble_fin} up to linear terms in initial conditions we find first cosmological time derivative of the Hubble function
\begin{equation}
\begin{split}
\frac{\dot{H}}{H_{0}^{2}} & \approx \frac{15}{8}\Omega_{\Lambda,0} - \\ & -\frac{69}{20}x^{*}\left(\Delta x +\frac{9}{9}\Delta z\right)\exp{\left(\frac{15}{4}\Omega_{\Lambda,0}\tau\right)}  \\ & - \frac{3}{160}x^{*}\left(16\Delta x+ 33\Delta z\right)\exp{\left(\frac{15}{8}\Omega_{\Lambda,0}\tau\right)}\,,
\end{split} 
\end{equation}
which at the critical point $\Omega_{\Lambda,0}\tau\to-\infty$ gives
\begin{equation}
\frac{\dot{H}}{H_{0}^{2}}\bigg|^{*} = \frac{15}{8}\Omega_{\Lambda,0}\,.
\end{equation}
The Hubble function vanishes in this limit at the critical point and we obtain that this state can correspond to a bouncing solution when $\Omega_{\Lambda,0}>0$ (Fig.~\ref{fig:6} bottom panel the point $B$) and to a collapsing solution when $\Omega_{\Lambda,0}<0$ (Fig.~\ref{fig:6} top panel the point $K$).

The next two critical points, denoted as points $d$ in Table \ref{tab:2}, with phase space coordinates
$$\left(x^{*}=\pm4\sqrt{-\ve\Omega_{\Lambda,0}}\,,z^{*}=\mp\frac{8}{3}\sqrt{-\ve\Omega_{\Lambda,0}}\right)$$
and the condition for the existence $-\ve\Omega_{\Lambda,0}>0$ are very interesting from the physical point of view. The energy conservation condition at this state is 
\begin{equation}
\left(\frac{H(a)}{H(a_{0})}\right)^{2}\Bigg|^{*} = \left(h^{*}\right)^{2} = \Omega_{\Lambda,0}\,,
\end{equation}
from which we obtain that this asymptotic state is in the physical region of the phase space only for the phantom scalar field $\ve=-1$ and positive cosmological constant $\Omega_{\Lambda,0}>0$. The eigenvalues of the linearisation matrix at this point are
\begin{equation}
\lambda_{1}=-3\Omega_{\Lambda,0}\,,\quad \lambda_{2}=3\Omega_{\Lambda,0}\,,
\end{equation}
of opposite signs and this gives that this state is in the form of a saddle type critical point. The linearised solutions are the following
\begin{equation}
\begin{split}
x(\tau) & = x^{*} -3\left(\Delta x + \frac{3}{2}\Delta z\right)\exp{(\lambda_{1}\tau)} \\ &+ 4\left(\Delta x + \frac{9}{8}\Delta z\right)\exp{(\lambda_{2}\tau)}\,,\\
z(\tau) & = z^{*} +\frac{8}{3}\left(\Delta x+\frac{3}{2}\Delta z\right)\exp{(\lambda_{1}\tau)}\\ &- \frac{8}{3}\left(\Delta x + \frac{9}{8}\Delta z\right)\exp{(\lambda_{2}\tau)}\,.
\end{split}
\end{equation}

From the dynamical time reparameterisation \eqref{eq:time_rep} we obtain differential equation for the scale factor as function of the time $\tau$
\begin{equation}
\ud\ln{a} \approx \ve\frac{1}{2}x^{*}\left(\Delta x + \frac{9}{8}\Delta z\right)\exp{(3\Omega_{\Lambda,0}\tau)}\ud\tau\,,
\end{equation}
end we should note that
along the stable direction with the eigenvalue $\lambda_{1}=-3\Omega_{\Lambda,0}$ and initial conditions $\Delta x + \frac{9}{8}\Delta z \equiv0$ there is no evolution of the scale factor thus there is evolution in the dynamical time $\tau$ given by the dynamics but there is no physical correspondence in the expansion of universe of this evolution. The Hubble function \eqref{eq:hubble_fin} up to terms in initial conditions is 
\begin{equation}
\left(\frac{H(a)}{H(a_{0})}\right)^{2} \approx \Omega_{\Lambda,0}+ \ve\frac{1}{2}x^{*}\left(\Delta x + \frac{3}{2}\Delta z\right)\exp{(\lambda_{1}\tau)}\,,
\end{equation}
and we read directly that for the initial conditions $\Delta x + \frac{3}{2}\Delta z \equiv0$ we obtain pure de Sitter type of evolution with constant value of the Hubble function.

Finally we have the following parametric solution for the Hubble function and the scale factor in the vicinity of the critical point
\begin{equation}
\left\{
\begin{split}
\left(\frac{H(a)}{H(a_{0})}\right)^{2} & \approx \Omega_{\Lambda,0}+ \ve\frac{1}{2}x^{*}\left(\Delta x + \frac{3}{2}\Delta z\right)\exp{(-3\Omega_{\Lambda,0}\tau)}\,,\\
\ln{\left(\frac{a}{a^{(i)}}\right)} & \approx \ve\frac{1}{6\Omega_{\Lambda,0}}x^{*}\left(\Delta x +\frac{9}{8}\Delta z\right)\big(\exp{(3\Omega_{\Lambda,0}\tau)}-1\big)\,.
\end{split}
\right.
\end{equation}
On the bottom panel of Fig.~\ref{fig:6} the pure de Sitter expansion is presented as the dotted trajectory going through the critical point under considerations. Note that the same type behaviour is also possible for the canonical scalar field (see top panel of Fig.~\ref{fig:5}). The global de Sitter evolution in the model under considerations do not distinguishes between the canonical and the phantom scalar field.

Now we proceed to the last critical point in the finite region of the phase space
$$
\left(x^{*}=\pm4\sqrt{-\ve\Omega_{\Lambda,0}}\,,z^{*}=\mp\frac{40}{9}\sqrt{-\ve\Omega_{\Lambda,0}}\right)$$
which is denoted as point $e$ in Table \ref{tab:2}.
The eigenvalues of the linearisation matrix calculated at this point gives
\begin{equation}
\lambda_{1}=-5\Omega_{\Lambda,0}\,, \quad \lambda_{2} = 5\Omega_{\Lambda,0}\,,
\end{equation}
which are real of opposite sings indicating for a saddle type critical point, and the linearised solutions to dynamics in the vicinity of this state are
\begin{equation}
\begin{split}
x(\tau) & = x^{*} + \frac{1}{10}\left(34\Delta x + 27\Delta z\right)\exp{(\lambda_{1}\tau)} \\ &- \frac{12}{5}\left(\Delta x + \frac{9}{8}\Delta z\right)\exp{(\lambda_{2}\tau)}\,,\\
z(\tau) & = z^{*} - \frac{4}{45}\left(34\Delta x + 27\Delta z\right)\exp{(\lambda_{1}\tau)} \\ & + \frac{136}{45}\left(\Delta x + \frac{9}{8}\Delta z\right)\exp{(\lambda_{2}\tau)}\,.
\end{split}
\end{equation}
Using these linearised solutions and from dynamical time reparameterisation \eqref{eq:time_rep} we obtain the following differential equation for the scale factor
\begin{equation}
\ud\ln{a} \approx -\ve\frac{1}{2}x^{*}\left(\Delta x + \frac{9}{8}\Delta z\right)\exp{(\lambda_{2}\tau)}\ud\tau\,.
\end{equation}
Then, solution of this equation together with a linearised approximation to the Hubble function \eqref{eq:hubble_fin} constitute parametric solution to the dynamics in the vicinity of the critical point
\begin{equation}
\left\{
\begin{split}
\left(\frac{H(a)}{H(a_{0})}\right)^{2} & \approx \frac{25}{9}\Omega_{\Lambda,0} \\ &-\ve\frac{1}{36}x^{*}\left(34\Delta x+ 27\Delta z\right)\exp{(-5\Omega_{\Lambda,0}\tau)} \\ &+\ve\frac{4}{9}\left(\Delta x + \frac{9}{8}\Delta z\right)\exp{(5\Omega_{\Lambda,0}\tau)}\,,\\
\ln{\left(\frac{a}{a^{(i)}}\right)} & \approx -\ve\frac{1}{10\Omega_{\Lambda,0}}x^{*}\left(\Delta x+\frac{9}{8}\Delta z\right)\big(\exp{(5\Omega_{\Lambda,0}\tau)}-1\big)\,.
\end{split}
\right.
\end{equation}

We have to note that the last two asymptotic states discussed here are very interesting both from mathematical and physical point of view. First, they are in form of a saddle critical points and it means that there is only non-generic set of initial conditions leading to these states, namely, a separatrix of a saddle. Second, a finite time is needed to achieve physical states represented by those points and there is nothing peculiar in physical behaviour at those points. A universe starting its evolution on either separatrix of the saddles reaches and crosses critical points in a finite amount of the scale factor. 

Now we are ready to analyse the system at infinity of the phase space. Using the projective coordinates \eqref{eq:proj_coor}
\begin{equation}
u\equiv\frac{x}{z}\,,\,\, \bar{w}\equiv\frac{1}{z^{2}}\,,
\end{equation}
we obtain the following dynamical system
\begin{equation}
\label{eq:const_infty_ln}
\begin{split}
\frac{\ud u}{\ud\ln{a}} & = -u(u+1)-\left(u+\frac{9}{8}\right)\left(\frac{\dot{H}}{H^{2}}+2\right)\,,\\
\frac{\ud \bar{w}}{\ud\ln{a}} & = -2\bar{w}\left(u+\frac{\dot{H}}{H^{2}}\right)\,,
\end{split}
\end{equation}
where the acceleration equation now reads
\begin{equation}
\label{eq:accel_inf}
\frac{\dot{H}}{H^{2}} = -2 + \frac{\ve\frac{1}{8}u^{2}+2\Omega_{\Lambda,0}\bar{w}}{\Omega_{\Lambda,0}\bar{w}+\ve\left(u+\frac{9}{8}\right)^{2}}\,,
\end{equation}
and the energy conservation condition is
\begin{equation}
\label{eq:hub_inf}
h^{2}=\left(\frac{H}{H_{0}}\right)^{2} = \Omega_{\Lambda,0}+\ve\frac{\left(u+\frac{3}{4}\right)\left(u+\frac{3}{2}\right)}{\bar{w}}\,.
\end{equation}

Using the following time reparameterisation
\begin{equation}
\frac{\ud}{\ud\eta}=\left(\Omega_{\Lambda,0}\bar{w}+\ve\left(u+\frac{9}{8}\right)^{2}\right) \frac{\ud}{\ud\ln{a}}
\end{equation}
we remove singularities in the denominator of the acceleration equation \eqref{eq:accel_inf} and dynamical system is given by
\begin{equation}
\label{eq:const_infty}
\begin{split}
\frac{\ud u}{\ud\eta} & = -\Omega_{\Lambda,0}\left(u+\frac{3}{2}\right)^{2}\bar{w}-\ve u\left(u+\frac{9}{8}\right)\left(u+\frac{3}{4}\right)\left(u+\frac{3}{2}\right)\,,\\
\frac{\ud \bar{w}}{\ud\eta} & = -2\bar{w}\left(\Omega_{\Lambda,0}u\bar{w}+\ve\left((u-2)\left(u+\frac{9}{8}\right)^{2}+\frac{1}{8}u^{2}\right)\right)\,,
\end{split}
\end{equation}
where the right hand sides of the equations are polynomial in dynamical variables.

\begin{table}
\caption{Critical points of the system \eqref{eq:const_infty} at infinity of the phase space defined as $\bar{w}=0$ together with values of the acceleration equation \eqref{eq:accel_inf}.}
\label{tab:3}
\renewcommand{\arraystretch}{1.5}
	\centering
	\begin{tabular}{|c|l|c|c|c|}
		\hline
		& $u^{*}$  & $\frac{\dot{H}}{H^{2}}\big|^{*}$ \\
		\hline
		R & $0$  & $-2$ \\
		S & $-\frac{9}{8}$ & $\pm\infty$  \\
		$\text{EdS}_{-}$ & $-\frac{3}{4}$ &  $-\frac{3}{2}$ \\
		$\text{dS}_{-}$ & $-\frac{3}{2}$ & $0$\\
		\hline
	\end{tabular}
\end{table}

In Table \ref{tab:3} we have gathered critical points of the system \eqref{eq:const_infty} at infinity of the phase space defined as $z\to\infty$ which correspond to $\bar{w}\equiv0$. Let us note that only the critical point denoted as $S$ will be investigated in the reparametrised time $\eta$ since only there denominator of the acceleration equation \eqref{eq:accel_inf} vanishes.

The critical point is located at
$$\left(u^{*}=-\frac{9}{8}\,,\quad \bar{w}^{*}=0\right)$$
and the eigenvalues of the linearisation matrix are
$$\lambda_{1}=-\ve\frac{81}{512}\,,\quad \lambda_{2}=-\ve\frac{81}{256}\,,$$
which indicates that for the phantom scalar field $\ve=-1$ it is an unstable node while for the canonical scalar field $\ve=+1$ is in the form of a stable node with respect to time $\eta$. The linearised solutions in the vicinity are the following
\begin{equation}
\begin{split}
u(\eta) & = u^{*} + \left(\Delta u -\ve\frac{8}{9}\Omega_{\Lambda,0}\Delta\bar{w}\right)\exp{(\lambda_{1}\eta)} \\ &+ \ve\frac{8}{9}\Omega_{\Lambda,0}\Delta\bar{w}\exp{(\lambda_{2}\eta)}\,,\\
\bar{w}(\eta) & = \Delta\bar{w}\exp{(\lambda_{2}\eta)}\,,
\end{split}
\end{equation}
where $\Delta u = u^{(i)} - u^{*}$ and $\Delta\bar{w}=\bar{w}^{(i)}$ are the initial conditions for the phase space variables.

Now, using these linearised solution and from \eqref{eq:hub_inf} we obtain the Hubble function up to linear terms in initial conditions
\begin{equation}
\begin{split}
\frac{H^{2}}{H_{0}^{2}} & \approx \Omega_{\Lambda,0}+\ve\frac{\left(\Delta u -\ve\frac{8}{9}\Omega_{\Lambda,0}\Delta\bar{w}\right)^{2}}{\Delta\bar{w}} \\ & -\ve\frac{9}{64}\frac{1}{\Delta\bar{w}}\exp{(-\lambda_{2}\eta)} \\ &+  \frac{16}{9}\Omega_{\Lambda,0}\left(\Delta u -\ve\frac{8}{9}\Omega_{\Lambda,0}\Delta\bar{w}\right)\exp{(\lambda_{1}\eta)} \\ & +\ve\frac{64}{81}\Omega_{\Lambda,0}^{2}\Delta\bar{w}\exp{(\lambda_{2}\eta)}\,,
\end{split}
\end{equation}
Lets consider only evolution toward the critical point $\lambda_{1}\eta$, $\lambda_{2}\eta \to -\infty$, then, starting form some time $\eta$ the third term starts to dominate and we obtain
\begin{equation}
\frac{H^{2}}{H_{0}^{2}} \approx -\ve\frac{9}{64}\frac{1}{\Delta\bar{w}}\exp{(-\lambda_{2}\eta)} >0\,,
\end{equation}
while $\Delta\bar{w}>0$ is always positive we obtain that $\ve=-1$, i.e. only for the phantom scalar field evolution takes place through the physical region of the phase space. From the energy conservation condition \eqref{eq:hub_inf} and from the acceleration equation \eqref{eq:accel_inf} we obtain that at this critical point both quantities blow up to infinty
$$
\frac{H^{2}}{H_{0}^{2}}\bigg|^{*} = \infty\,, \quad \frac{\dot{H}}{H^{2}}\bigg|^{*} = \pm\infty\,.
$$

Now, we can show that this singularity is of a finite scale factor type. The differential equation for the scale factor up to linear terms in initial conditions for the phantom scalar field $\ve=-1$ is in the following form
$$
\ud\ln{a} = \Omega_{\Lambda,0}\Delta\bar{w}\exp{\left(\frac{81}{256}\eta\right)}\ud\eta\,,
$$
from which we read that for $\Omega_{\Lambda,0}>0$ when time $\eta$ grows so the scale factor $a$ grows. Taking the evolution toward critical point we can integrate this equation
$$
\int_{a^{(i)}}^{a^{(f)}}\ud\ln{a} = \Omega_{\Lambda,0}\Delta\bar{w}\int_{0}^{-\infty}\exp{\left(\frac{81}{256}\eta\right)}\ud\eta\,,
$$
where initial value of the scale factor $a^{(i)}$ corresponds to initial value of time $\eta=0$ and final value of the scale factor $a^{(f)}$ corresponds to value at the asymptotic state at $\eta=-\infty$. We obtain
$$
\ln{\left(\frac{a^{(f)}}{a^{(i)}}\right)} = - \frac{256}{81}\Omega_{\Lambda,0}\Delta\bar{w}\,,
$$
where for
$$
\Omega_{\Lambda,0}>0:\qquad \ln{\left(\frac{a^{(f)}}{a^{(i)}}\right)} <0\quad \Rightarrow \quad 0<a^{(f)}<a^{(i)}\,,
$$
which gives that a finite scale factor singularity is unstable with respect to the expansion of the universe.

We can also show that despite this critical point is located at infinity of the phase space it corresponds to a finite value of the scalar field $\phi$. From definition of the variable $\bar{w}$ we recover the scalar field
$$
\frac{\kappa^{2}}{6}\phi^{2} = \frac{1}{\bar{w}\frac{H^{2}}{H_{0}^{2}}}\,,
$$
then from the linearised solutions and the Hubble function we obtain that value of the scalar field at the critical point is
$$
\frac{\kappa^{2}}{6}\phi^{2}\bigg|^{*} =\lim_{\eta\to-\infty}\frac{\kappa^{2}}{6}\phi^{2} = \frac{64}{9}\,.
$$

Now we are ready to investigate behaviour of the remaining asymptotic states form Table \ref{tab:3} studying the dynamical system \eqref{eq:const_infty_ln} where the evolution is in terms of natural logarithm of the scale factor, namely, expansion of the universe.

The first critical point is
$$\left(u^{*}=0\,,\quad \bar{w}^{*}=0\right)$$
the acceleration equation calculated at this point gives 
$$
\frac{\dot{H}}{H^{2}}\bigg|^{*} = -2\,,
$$
which corresponds to a ``radiation'' dominated universe and dynamical behaviour in the vicinity of this point mimics radiation dominated expansion even that there is no radiation-like form of matter included in the model.

The eigenvalues of the linearisation matrix are
$$\lambda_{1}=-1\,,\quad \lambda_{2}=4\,,$$
which points to a saddle type critical point and the linearised solutions are the following
\begin{equation}
\begin{split}
u(a) & = \left(\Delta u +\ve\frac{16}{45}\Omega_{\Lambda,0}\Delta\bar{w}\right)\left(\frac{a}{a^{(i)}}\right)^{-1} \\&- \ve\frac{16}{45}\Omega_{\Lambda,0}\Delta\bar{w}\left(\frac{a}{a^{(i)}}\right)^{4}\,,\\
\bar{w}(a) & = \Delta\bar{w}\left(\frac{a}{a^{(i)}}\right)^{4}\,,
\end{split}
\end{equation}
where $\Delta u = u^{(i)}$, $\Delta\bar{w} = \bar{w}^{(i)}$ are initial conditions for the phase space variables and $a^{(i)}$ is initial value of the scale factor.

The Hubble function \eqref{eq:hub_inf} up to linear terms in initial conditions is 
\begin{equation}
\begin{split}
\left(\frac{H(a)}{H(a_{0})}\right)^{2}  \approx & \,\,\ve\frac{9}{8}\frac{1}{\Delta\bar{w}}\left(\frac{a}{a^{(i)}}\right)^{-4}+ \frac{1}{5}\Omega_{\Lambda,0}   \\& + \ve\frac{9}{4}\left(\frac{\Delta u}{\Delta\bar{w}}+\ve\frac{16}{45}\Omega_{\Lambda,0}\right)\left(\frac{a}{a^{(i)}}\right)^{-5}  \\ & + \ve\frac{1}{\Delta\bar{w}}\left(\Delta u + \ve\frac{16}{45}\Omega_{\Lambda,0}\Delta\bar{w}\right)^{2}\left(\frac{a}{a^{(i)}}\right)^{-6}  \\ &+\ve\left(\frac{16}{45}\Omega_{\Lambda,0}\right)^{2}\Delta\bar{w}\left(\frac{a}{a^{(i)}}\right)^{4}  \\ & - \frac{32}{45}\Omega_{\Lambda,0}\left(\Delta u + \ve\frac{16}{45}\Omega_{\Lambda,0}\Delta\bar{w}\right)\left(\frac{a}{a^{(i)}}\right)^{-1}\,,
\end{split}
\end{equation}
where for $\Delta\bar{w}<<1$ the first term dominates during evolution in the vicinity of this state giving rise to a radiation-like behaviour of the model
\begin{equation}
\left(\frac{H(a)}{H(a_{0})}\right)^{2} \approx \ve\frac{9}{8}\frac{1}{\Delta\bar{w}}\left(\frac{a}{a^{(i)}}\right)^{-4}\,.
\end{equation}
Since $\Delta\bar{w}>0$ is always positive we have that the evolution in the vicinity of this point takes place in physical region of the phase space only for the canonical scalar field $\ve=+1$. Next, as for the previous critical point, we can also recover value of the scalar field. Using linearised solutions we find that  
\begin{equation}
\frac{\kappa^{2}}{6}(\phi^{*})^{2} = \ve\frac{8}{9},
\end{equation}
and we obtain that $\ve=+1$ and even the critical point is located at infinity of the phase space it corresponds to finite value of the scalar field.

The next critical point located at
$$\left(u^{*}=-\frac{3}{4}\,,\quad \bar{w}^{*}=0\right)$$
with the acceleration equation \eqref{eq:accel_inf} 
$$ \frac{\dot{H}}{H^{2}}\bigg|^{*} = -\frac{3}{2}\,,$$
which points toward Einstein-de Sitter asymptotic state. The eigenvalues of the linearisation matrix are
$$\lambda_{1}=\frac{3}{2}\,, \quad \lambda_{2}=\frac{9}{2}\,,$$
and this state is in the form of an unstable node with respect to expansion of the universe. The linearised solutions are
\begin{equation}
\begin{split}
u(a) & = u^{*}+ \left(\Delta u +\ve\frac{4}{3}\Omega_{\Lambda,0}\Delta\bar{w}\right)
\left(\frac{a}{a^{(i)}}\right)^{\frac{3}{2}}  \\ & - \ve\frac{4}{3}\Omega_{\Lambda,0}\Delta\bar{w}
\left(\frac{a}{a^{(i)}}\right)^{\frac{9}{2}}\,,\\
\bar{w}(a) & = \Delta\bar{w}\left(\frac{a}{a^{(i)}}\right)^{\frac{9}{2}}\,,
\end{split}
\end{equation}
where $\Delta u = u^{(i)}-u^{*}$, $\Delta\bar{w} = \bar{w}^{(i)}$ are initial conditions for the phase space variables and $a^{(i)}$ is initial value of the scale factor.

Up to linear terms in initial conditions the Hubble function \eqref{eq:hub_inf} is 
\begin{equation}
\begin{split}
\left(\frac{H(a)}{H(a_{0})}\right)^{2} & \approx \left(\Omega_{\Lambda,0}+\ve\frac{3}{4}\frac{\Delta u}{\Delta\bar{w}}\right)\left(\frac{a}{a^{(i)}}\right)^{-3}  \\ & +\ve\frac{1}{\Delta\bar{w}}\left(\Delta u +\ve\frac{4}{3}\Omega_{\Lambda,0}\Delta\bar{w}\right)^{2}\left(\frac{a}{a^{(i)}}\right)^{-\frac{3}{2}}  \\ & + \ve\frac{16}{9}\Omega_{\Lambda,0}^{2}\Delta\bar{w}\left(\frac{a}{a^{(i)}}\right)^{\frac{9}{2}}  \\ & - \frac{8}{3}\Omega_{\Lambda,0}\left(\Delta u + \ve\frac{4}{3} \Omega_{\Lambda,0}\Delta\bar{w}\right)\left(\frac{a}{a^{(i)}}\right)^{\frac{3}{2}}\,,
\end{split}
\end{equation}
where we observe that the first term is of the zeroth order in initial conditions and dominates during the linearised evolution. Taking a limit $a\to0$ to the asymptotic state we also observe that the first term dominates leading to the following approximation to the Hubble functions
\begin{equation}
\left(\frac{H(a)}{H(a_{0})}\right)^{2} \approx \left(\Omega_{\Lambda,0}+\ve\frac{3}{4}\frac{\Delta u}{\Delta\bar{w}} \right)\left(\frac{a}{a^{(i)}}\right)^{-3}\,,
\end{equation}
which exactly corresponds to Einstein-de Sitter type of evolution. In order to obtain the evolution in physical region of the phase space the following initial conditions must be met
$$
\ve\Delta u > -\frac{4}{3}\Omega_{\Lambda,0}\Delta\bar{w}\,.
$$

The final critical point located at
$$ \left( u^{*} = -\frac{3}{2}\,, \quad \bar{w}^{*} =0 \right) $$
with the vanishing acceleration equation \eqref{eq:accel_inf} describes the de Sitter state. The eigenvalues of the linearisation matrix calculated at this point are
$$ \lambda_{1}=\lambda_{2}=3\,, $$
which indicates that the asymptotic state is unstable during expansion. The linearised solutions are
\begin{equation}
\begin{split}
u(a) & = u^{*} + \Delta u \left(\frac{a}{a^{(i)}}\right)^{3}\,,\\
\bar{w}(a) & = \Delta\bar{w}\left(\frac{a}{a^{(i)}}\right)^{3}\,,
\end{split}
\end{equation}
where $\Delta u = u^{(i)} - u^{*}$ and $\Delta \bar{w} = \bar{w}^{(i)}$ are the initial conditions in the vicinity of the state. The Hubble function \eqref{eq:hub_inf} can be approximated as
\begin{equation}
\left(\frac{H(a)}{H(a_{0})}\right)^{2} \approx \Omega_{\Lambda,0} -\ve\frac{3}{4}\frac{\Delta u}{\Delta\bar{w}} + \ve \frac{(\Delta u)^{2}}{\Delta\bar{w}}\left(\frac{a}{a^{(i)}}\right)^{3}\,,
\end{equation}
and taking the limit $a\to0$ we obtain the value of the Hubble function at the asymptotic state
\begin{equation}
\left(\frac{H(0)}{H(a_{0})}\right)^{2} \approx \Omega_{\Lambda,0}-\ve\frac{3}{4}\frac{\Delta u}{\Delta\bar{w}} = \text{const.}>0\,.
\end{equation}
First, we have to note that the above quantity must be positive in order to obtain a physical state, as $\Delta\bar{w}>0$, for fixed value of $\Omega_{\Lambda,0}$ one have to suitably choose $\ve\Delta u < \frac{4}{3}\Omega_{\Lambda,0}\Delta\bar{w}$ and this is possible even for $\Omega_{\Lambda,0}\equiv0$ or $\Omega_{\Lambda,0}<0$. Second, every trajectory with different initial conditions lead to different values of the Hubble function at the asymptotic state, i.e. different values of the effective energy density associated with this state
\begin{equation}
\left(\frac{H(0)}{H(a_{0})}\right)^{2} = \frac{\kappa^{2}\rho_{\text{eff}}}{3H_{0}^{2}}\Big|^{*} \approx \Omega_{\Lambda,0} -\ve\frac{3}{4}\frac{\Delta u}{\Delta\bar{w}}\,,
\end{equation}
additionally there is an open and dense set of initial condition for which energy density at de Sitter state is smaller then the Planck energy density
\begin{equation}
\Omega_{\Lambda,0} -\ve\frac{3}{4}\frac{\Delta u}{\Delta\bar{w}} < \frac{8\pi}{3}\frac{m_{Pl}^{2}}{H_{0}^{2}}\,.
\end{equation}
The Hubble function in the vicinity of the de Sitter can be presented as
%\begin{widetext}
\begin{equation}
\begin{split}
\left(\frac{H(a)}{H(a_{0})}\right)^{2} & \approx \left(\frac{H(0)}{H(a_{0})}\right)^{2}  \\ & +\left(\left(\frac{H(a^{(i)})}{H(a_{0})}\right)^{2}-\left(\frac{H(0)}{H(a_{0})}\right)^{2}\right)\left(\frac{a}{a^{(i)}}\right)^{3}\,,
\end{split}
\end{equation}
%\end{widetext}
\noindent
where the initial value of the Hubble function is
\begin{equation}
\begin{split}
\left(\frac{H(a^{(i)})}{H(a_{0})}\right)^{2} & \approx \Omega_{\Lambda,0} -\ve\frac{3}{4}\frac{\Delta u}{\Delta\bar{w}} + \ve \frac{(\Delta u)^{2}}{\Delta\bar{w}} \\ & \approx \left(\frac{H(0)}{H(a_{0})}\right)^{2} + \ve \frac{(\Delta u)^{2}}{\Delta\bar{w}}\,,
\end{split}
\end{equation}
and again as at the end of the previous section one can conclude that since $\Delta\bar{w}$ is always positive, the energy density at $a=a^{(i)}>0$ can be larger than the energy density of the de Sitter state at $a=0$.

On Figs.~\ref{fig:5} and \ref{fig:6} we presented four generic cases for the structure of the phase space and dynamical behaviour of the model under considerations. On the top panel on Fig.~\ref{fig:5} we have phase space portrait of the model with the canonical scalar field $\ve=+1$ and positive value of the cosmological constant $\Omega_{\Lambda,0}>0$. We observe that there there is a open and dense set of initial conditions giving rise to non-singular cosmological evolution from an unstable de Sitter state toward a stable one. Additionally there is also an open and dense set of initial conditions giving cosmological evolution from an unstable Einstein-de Sitter state toward a stable de Sitter. On the bottom panel on Fig.~\ref{fig:5} we have phase space diagram for the model with vanishing cosmological constant $\Omega_{\Lambda,0}=0$. The critical point $D$ corresponds to a degenerated critical point with two stable sectors and four saddle sectors of the phase space and its structure can be understood as merger of critical points from the previous diagram. We conclude that not only non-minimal coupling constant can be treated as a bifurcation parameter but also the value of globally constant scalar field potential function \cite{Humieja:2019ywy}. On Fig.~\ref{fig:6} top panel we presented dynamical behaviour for the model with canonical scalar field $\ve=+1$ and the negative cosmological constant $\Omega_{\Lambda,0}<0$. In the physical region of the phase space there is to possible initial states for the universe, namely, initial non-singular de Sitter state and Einstein-de Sitter state. In this case both types of evolution lead to the collapsing solution $K$ for finite value of the scale factor. On the bottom panel of Fig.~\ref{fig:6} we presented the phase space diagram for the model with phantom scalar field $\ve=-1$ and positive cosmological constant $\Omega_{\Lambda,0}>0$. It is very interesting that in this case there is only single trajectory connecting initial de Sitter state $deS_{-}$ with the final de Sitter state $deS_{+}$. This trajectory corresponds to pure de Sitter evolution of the universe. 

\section{Vanishing potential function and the exact solutions}

The system \eqref{eq:const_infty} for the vanishing potential function of the scalar field $\Omega_{\Lambda,0}=0$ and the natural logarithm of the scale factor as a time variable takes the following form
\begin{equation}
\begin{split}
\label{eq:sys_zero}
(9+8u)\frac{\ud u}{\ud\ln{a}} & = -u(3+2u)(3+4u)\,,\\
(9+8u)^{2}\frac{\ud \bar{w}}{\ud\ln{a}} & = -2\bar{w}\big((u-2)(9+8u)^{2}+8u^{2}\big)\,.
\end{split}
\end{equation}
First, we should note that the first equation in this system is decoupled from the second, i.e. finding the solution to this equation we essentially can solve the second equation. Second, the equation constitutes special form of Abel ordinary differential equation of the second kind which can be converted to an equation of the first kind using a suitable coordinate transformation. We will follow different route in order to find the exact solution to the system. The change of time parameter $\ln{a}\to \frac{1}{3}\ln{a^{3}}$ gives the following equation
$$
\frac{1}{a^{3}}\ud a^{3} = -3\frac{9+8u}{u(3+2u)(3+4u)}\ud u\,,
$$
after integration we obtain the implicit form of the solution
\begin{equation}
\label{eq:u_a}
\left(\frac{a}{a^{(i)}}\right)^{3} = \left(\frac{u^{(i)}}{u}\right)^{3}\frac{(3+2u)(3+4u)^{2}}{(3+2u^{(i)})(3+4u^{(i)})^{2}}\,,
\end{equation}
where $a^{(i)}$ and $u^{(i)}$ are the initial conditions and one can solve this equation in order to obtain $u=u(a)$.
To solve the second equation of the system \eqref{eq:sys_zero} we assume that the $\bar{w}$ dynamical variable is a function of the $u$ phase space variable
$$
\bar{w}=\bar{w}(u)\,.
$$
Then we obtain
$$
\frac{1}{\bar{w}}\ud\bar{w} = 2 \frac{(u-2)(9+8u)^{2}+8 u^{2}}{u(3+2u)(3+4u)(9+8u)}\ud u\,,
$$
and integration gives
\begin{equation}
\label{eq:w_u}
\frac{\bar{w}}{\bar{w}^{(i)}} = \left(\frac{u^{(i)}}{u}\right)^{4}\frac{(3+2u)(3+4u)^{3}(9+8u)^{2}}{(3+2u^{(i)})(3+4u^{(i)})^{3}(9+8u^{(i)})^{2}}\,,
\end{equation}
where $\bar{w}^{(i)}$ and $u^{(i)}$ are the initial conditions. Next, from the energy conservation condition \eqref{eq:hub_inf} we have that
$$
\left(\frac{H(a^{(i)})}{H(a_{0})}\right)^{2} = \ve\frac{1}{8}\frac{(3+2u^{(i)})(3+4u^{(i)})}{\bar{w}^{(i)}} > 0\,,
$$
and since the $\bar{w}$ variable is always positive the condition for the evolution in the physical region of the phase space is
\begin{equation}
\ve(3+2u^{(i)})(3+4u^{(i)})>0\,.
\end{equation}
Finally, having the function $u=u(a)$ from \eqref{eq:u_a} one can obtain $\bar{w}=\bar{w}(a)$ from \eqref{eq:w_u} and from the energy conservation condition \eqref{eq:hub_inf} one arrives at the exact form of the Hubble function
\begin{equation}
\left(\frac{H(a)}{H(a_{0})}\right)^{2} = \ve\frac{1}{8}\frac{\big(3+2u(a)\big)\big(3+4u(a)\big)}{\bar{w}(a)}\,.
\end{equation}

\section{Conclusions}

In this paper we have investigated dynamics of a flat FRW cosmological model filled with the non-minimally coupled scalar field with a potential function. With assumption of the monomial type of behaviour at infinite values of the scalar field we were able to find the specific value of the non-minimal coupling constant $\xi=\frac{3}{16}$ for which there were de Sitter and Einstein-de Sitter states. Using dynamical systems methods we were able to constrain the slope of the potential function at infinity for which the asymptotic de Sitter and Einstein-de Sitter states are in the form of unstable critical points with respect to expansion of the universe.   
We have found that the asymptotic unstable with respect to expansion of the universe de Sitter and Einstein-de Sitter states exist both for negative and positive potential functions. For monomial scalar field potential functions $U(\phi)\propto U_{0}\phi^{\alpha}$ with $\alpha<1$ both the Einstein-de Sitter and the de Sitter states are unstable with respect to the expansion of the universe (Figs.~\ref{fig:1} and \ref{fig:2}); for $1<\alpha<5$ the Einstein-de Sitters state corresponds to an unstable node while the de Sitter state is in the form of a saddle critical point (Fig.~\ref{fig:3}); for $\alpha>5$ both state are in the form of saddle type critical points (Fig.~\ref{fig:4}). We have found additional interesting feature that for a negative potential function there is a possibility for an asymptotically stable de Sitter state (see Fig.~\ref{fig:4}).

Global dynamical analysis of model with a constant potential function and the non-minimal coupling constant $\xi=\frac{3}{16}$ was performed. We were able to show that the asymptotically unstable de Sitter state and asymptotically unstable Einstein-de Sitter state exist. For the positive cosmological constant there is an open and dense set of initial conditions giving rise to non-singular evolution of the universe from an unstable de Sitter state toward a stable one.

The obtained possible evolutional path can indicate for some generalisations of the seminal Starobinsky type of evolution \cite{Starobinsky:1980te}, as well as obtained evolution was not carefully designated \cite{Mukhanov:1991zn, Brandenberger:1993ef}. The value of the non-minimal coupling constant $\xi=\frac{3}{16}$ corresponds to conformal coupling value in a $5-$dimensional theory of gravity. Presented analysis and results might point toward a new fundamental symmetry in the matter sector of the theory \cite{tHooft:2014daa,Hrycyna:2017oug}.

\begin{acknowledgements}
I am grateful to Marek Szyd{\l}owski for valuable discussions and comments. 
\end{acknowledgements}

\bibliographystyle{hspphys}
\bibliography{../bib/confinv,../bib/darkenergy,../bib/quintessence,../bib/quartessence,../bib/astro,../bib/dynamics,../bib/standard,../bib/inflation,../bib/sm_nmc,../bib/singularities,../bib/JvsE,../bib/moje}

\begin{thebibliography}{10}
\providecommand{\url}[1]{{#1}}
\providecommand{\urlprefix}{URL }
\expandafter\ifx\csname urlstyle\endcsname\relax
  \providecommand{\doi}[1]{DOI \discretionary{}{}{}#1}\else
  \providecommand{\doi}{DOI \discretionary{}{}{}\begingroup
  \urlstyle{rm}\Url}\fi

\bibitem{Riess:1998cb}
{\bf Supernova Search Team} Collaboration, A.G. Riess, A.V. Filippenko,
  P.~Challis, A.~Clocchiattia, A.~Diercks, P.M. Garnavich, R.L. Gilliland, C.J.
  Hogan, S.~Jha, R.P. Kirshner, B.~Leibundgut, M.M. Phillips, D.~Reiss, B.P.
  Schmidt, R.A. Schommer, R.C. Smith, J.~Spyromilio, C.~Stubbs, N.B. Suntzeff,
  J.~Tonry, Astron. J. \textbf{116}, 1009 (1998).
\newblock \doi{10.1086/300499}.
\newblock [\href{http://arxiv.org/abs/astro-ph/9805201}{{\tt
  astro-ph/9805201}}]

\bibitem{Perlmutter:1998np}
{\bf Supernova Cosmology Project} Collaboration, S.~Perlmutter, G.~Aldering,
  G.~Goldhaber, R.~Knop, P.~Nugent, P.~Castro, S.~Deustua, S.~Fabbro,
  A.~Goobar, D.~Groom, I.M. Hook, A.~Kim, M.~Kim, J.~Lee, N.~Nunes, C.P.
  R.~Pain, R.~Quimby, C.~Lidman, R.~Ellis, M.~Irwin, R.~McMahon,
  P.~Ruiz-Lapuente, N.~Walton, B.~Schaefer, B.~Boyle, A.~Filippenko,
  T.~Matheson, A.~Fruchter, N.~Panagia, H.~Newberg, W.~Couch, Astrophys. J.
  \textbf{517}, 565 (1999).
\newblock \doi{10.1086/307221}.
\newblock [\href{http://arxiv.org/abs/astro-ph/9812133}{{\tt
  astro-ph/9812133}}]

\bibitem{Copeland:2006wr}
E.J. Copeland, M.~Sami, S.~Tsujikawa, Int.~J.~Mod.~Phys. \textbf{D15}, 1753
  (2006).
\newblock \doi{10.1142/S021827180600942X}.
\newblock [\href{http://arxiv.org/abs/hep-th/0603057}{{\tt hep-th/0603057}}]

\bibitem{Bahamonde:2017ize}
S.~Bahamonde, C.G. Böhmer, S.~Carloni, E.J. Copeland, W.~Fang, N.~Tamanini,
  Phys. Rept. \textbf{775-777}, 1 (2018).
\newblock \doi{10.1016/j.physrep.2018.09.001}.
\newblock [\href{http://arxiv.org/abs/1712.03107}{{\tt arXiv:1712.03107}}]

\bibitem{Ratra:1987rm}
B.~Ratra, P.J.E. Peebles, Phys.~Rev. \textbf{D37}, 3406 (1988).
\newblock \doi{10.1103/PhysRevD.37.3406}

\bibitem{Wetterich:1987fm}
C.~Wetterich, Nucl.~Phys. \textbf{B302}, 668 (1988).
\newblock \doi{10.1016/0550-3213(88)90193-9}.
\newblock [\href{http://arxiv.org/abs/1711.03844}{{\tt arXiv:1711.03844}}]

\bibitem{Chernikov:1968zm}
N.A. Chernikov, E.A. Tagirov, Annales Poincare Phys. Theor. \textbf{A9}, 109
  (1968)

\bibitem{Callan:1970ze}
C.G. Callan, Jr., S.R. Coleman, R.~Jackiw, Annals Phys. \textbf{59}, 42 (1970).
\newblock \doi{10.1016/0003-4916(70)90394-5}

\bibitem{Birrell:1979ip}
N.D. Birrell, P.C.W. Davies, Phys. Rev. \textbf{D22}, 322 (1980).
\newblock \doi{10.1103/PhysRevD.22.322}

\bibitem{Donoghue:1994dn}
J.F. Donoghue, Phys.Rev. \textbf{D50}, 3874 (1994).
\newblock \doi{10.1103/PhysRevD.50.3874}.
\newblock [\href{http://arxiv.org/abs/gr-qc/9405057}{{\tt gr-qc/9405057}}]

\bibitem{Allen:1983dg}
B.~Allen, Nucl.~Phys. \textbf{B226}, 228 (1983).
\newblock \doi{10.1016/0550-3213(83)90470-4}

\bibitem{Ishikawa:1983kz}
K.~Ishikawa, Phys.~Rev. \textbf{D28}, 2445 (1983).
\newblock \doi{10.1103/PhysRevD.28.2445}

\bibitem{Birrell:1984ix}
N.D. Birrell, P.C.W. Davies, \emph{Quantum Fields in Curved Space} (Cambridge
  University Press, Cambridge, 1984)

\bibitem{Parker:book}
L.E. Parker, D.J. Toms, \emph{{Quantum Field Theory in Curved Spacetime.
  Quantized Fields and Gravity}} (Cambridge University Press, Cambridge, 2009)

\bibitem{Maeda:1985bq}
K.i. Maeda, Class. Quant. Grav. \textbf{3}, 233 (1986).
\newblock \doi{10.1088/0264-9381/3/2/017}

\bibitem{Accetta:1985du}
F.S. Accetta, D.J. Zoller, M.S. Turner, Phys. Rev. \textbf{D31}, 3046 (1985).
\newblock \doi{10.1103/PhysRevD.31.3046}

\bibitem{Planck:2013jfk}
{\bf Planck} Collaboration, P.~Ade, et~al., Astron.~Astrophys. \textbf{571},
  A22 (2014).
\newblock \doi{10.1051/0004-6361/201321569}.
\newblock [\href{http://arxiv.org/abs/1303.5082}{{\tt arXiv:1303.5082}}]

\bibitem{Martin:2013nzq}
J.~Martin, C.~Ringeval, R.~Trotta, V.~Vennin, JCAP \textbf{03}, 039 (2014).
\newblock \doi{10.1088/1475-7516/2014/03/039}.
\newblock [\href{http://arxiv.org/abs/1312.3529}{{\tt arXiv:1312.3529}}]

\bibitem{Ade:2015lrj}
{\bf Planck} Collaboration, P.A.R. Ade, et~al., Astron. Astrophys.
  \textbf{594}, A20 (2016).
\newblock \doi{10.1051/0004-6361/201525898}.
\newblock [\href{http://arxiv.org/abs/1502.02114}{{\tt arXiv:1502.02114}}]

\bibitem{Kobayashi:2011nu}
T.~Kobayashi, M.~Yamaguchi, J.~Yokoyama, Prog.~Theor.~Phys. \textbf{126}, 511
  (2011).
\newblock \doi{10.1143/PTP.126.511}.
\newblock [\href{http://arxiv.org/abs/1105.5723}{{\tt arXiv:1105.5723}}]

\bibitem{Atkins:2010eq}
M.~Atkins, X.~Calmet, Phys. Lett. \textbf{B695}, 298 (2011).
\newblock \doi{10.1016/j.physletb.2010.10.049}.
\newblock [\href{http://arxiv.org/abs/1002.0003}{{\tt arXiv:1002.0003}}]

\bibitem{Atkins:2010re}
M.~Atkins, X.~Calmet, Eur. Phys. J. \textbf{C70}, 381 (2010).
\newblock \doi{10.1140/epjc/s10052-010-1476-2}.
\newblock [\href{http://arxiv.org/abs/1005.1075}{{\tt arXiv:1005.1075}}]

\bibitem{Luo:2005ra}
M.X. Luo, Q.P. Su, Phys.~Lett. \textbf{B626}, 7 (2005).
\newblock \doi{10.1016/j.physletb.2005.08.050}.
\newblock [\href{http://arxiv.org/abs/astro-ph/0506093}{{\tt
  astro-ph/0506093}}]

\bibitem{Nozari:2007eq}
K.~Nozari, S.D. Sadatian, Mod.~Phys.~Lett. \textbf{A23}, 2933 (2008).
\newblock \doi{10.1142/S0217732308026698}.
\newblock [\href{http://arxiv.org/abs/0710.0058}{{\tt arXiv:0710.0058}}]

\bibitem{Szydlowski:2008zza}
M.~Szydlowski, O.~Hrycyna, A.~Kurek, Phys. Rev. \textbf{D77}, 027302 (2008).
\newblock \doi{10.1103/PhysRevD.77.027302}.
\newblock [\href{http://arxiv.org/abs/0710.0366}{{\tt arXiv:0710.0366}}]

\bibitem{Atkins:2012yn}
M.~Atkins, X.~Calmet, Phys.~Rev.~Lett. \textbf{110}, 051301 (2013).
\newblock \doi{10.1103/PhysRevLett.110.051301}.
\newblock [\href{http://arxiv.org/abs/1211.0281}{{\tt arXiv:1211.0281}}]

\bibitem{Hrycyna:2015vvs}
O.~Hrycyna, Phys. Lett. \textbf{B768}, 218 (2017).
\newblock \doi{10.1016/j.physletb.2017.02.062}.
\newblock [\href{http://arxiv.org/abs/1511.08736}{{\tt arXiv:1511.08736}}]

\bibitem{Spokoiny:1984bd}
B.L. Spokoiny, Phys.~Lett. \textbf{B147}, 39 (1984).
\newblock \doi{10.1016/0370-2693(84)90587-2}

\bibitem{Ford:1986sy}
L.~Ford, Phys. Rev. \textbf{D35}, 2955 (1987).
\newblock \doi{10.1103/PhysRevD.35.2955}

\bibitem{Salopek:1988qh}
D.S. Salopek, J.R. Bond, J.M. Bardeen, Phys.~Rev. \textbf{D40}, 1753 (1989).
\newblock \doi{10.1103/PhysRevD.40.1753}

\bibitem{Amendola:1990nn}
L.~Amendola, M.~Litterio, F.~Occhionero, Int.~J.~Mod.~Phys. \textbf{A5}, 3861
  (1990).
\newblock \doi{10.1142/S0217751X90001653}

\bibitem{Fakir:1992cg}
R.~Fakir, S.~Habib, W.~Unruh, Astrophys.~J. \textbf{394}, 396 (1992).
\newblock \doi{10.1086/171591}

\bibitem{Barvinsky:1994hx}
A.O. Barvinsky, A.Y. Kamenshchik, Phys.~Lett. \textbf{B332}, 270 (1994).
\newblock \doi{10.1016/0370-2693(94)91253-X}.
\newblock [\href{http://arxiv.org/abs/gr-qc/9404062}{{\tt gr-qc/9404062}}]

\bibitem{Faraoni:1996rf}
V.~Faraoni, Phys.~Rev. \textbf{D53}, 6813 (1996).
\newblock \doi{10.1103/PhysRevD.53.6813}.
\newblock [\href{http://arxiv.org/abs/astro-ph/9602111}{{\tt
  astro-ph/9602111}}]

\bibitem{Barvinsky:1998rn}
A.O. Barvinsky, A.Y. Kamenshchik, Nucl.~Phys. \textbf{B532}, 339 (1998).
\newblock \doi{10.1016/S0550-3213(98)00484-2}.
\newblock [\href{http://arxiv.org/abs/hep-th/9803052}{{\tt hep-th/9803052}}]

\bibitem{Barvinsky:2008ia}
A.O. Barvinsky, A.Y. Kamenshchik, A.A. Starobinsky, JCAP \textbf{11}, 021
  (2008).
\newblock \doi{10.1088/1475-7516/2008/11/021}.
\newblock [\href{http://arxiv.org/abs/0809.2104}{{\tt arXiv:0809.2104}}]

\bibitem{Setare:2008mb}
M.R. Setare, E.N. Saridakis, JCAP \textbf{03}, 002 (2009).
\newblock \doi{10.1088/1475-7516/2009/03/002}.
\newblock [\href{http://arxiv.org/abs/0811.4253}{{\tt arXiv:0811.4253}}]

\bibitem{Setare:2008pc}
M.R. Setare, E.N. Saridakis, Phys.~Lett. \textbf{B671}, 331 (2009).
\newblock \doi{10.1016/j.physletb.2008.12.026}.
\newblock [\href{http://arxiv.org/abs/0810.0645}{{\tt arXiv:0810.0645}}]

\bibitem{Uzan:1999ch}
J.P. Uzan, Phys.~Rev. \textbf{D59}, 123510 (1999).
\newblock \doi{10.1103/PhysRevD.59.123510}.
\newblock [\href{http://arxiv.org/abs/gr-qc/9903004}{{\tt gr-qc/9903004}}]

\bibitem{Chiba:1999wt}
T.~Chiba, Phys.~Rev. \textbf{D60}, 083508 (1999).
\newblock \doi{10.1103/PhysRevD.60.083508}.
\newblock [\href{http://arxiv.org/abs/gr-qc/9903094}{{\tt gr-qc/9903094}}]

\bibitem{Amendola:1999qq}
L.~Amendola, Phys.~Rev. \textbf{D60}, 043501 (1999).
\newblock \doi{10.1103/PhysRevD.60.043501}.
\newblock [\href{http://arxiv.org/abs/astro-ph/9904120}{{\tt
  astro-ph/9904120}}]

\bibitem{Holden:1999hm}
D.J. Holden, D.~Wands, Phys.~Rev. \textbf{D61}, 043506 (2000).
\newblock \doi{10.1103/PhysRevD.61.043506}.
\newblock [\href{http://arxiv.org/abs/gr-qc/9908026}{{\tt gr-qc/9908026}}]

\bibitem{Bartolo:1999sq}
N.~Bartolo, M.~Pietroni, Phys.~Rev. \textbf{D61}, 023518 (2000).
\newblock \doi{10.1103/PhysRevD.61.023518}.
\newblock [\href{http://arxiv.org/abs/hep-ph/9908521}{{\tt hep-ph/9908521}}]

\bibitem{Boisseau:2000pr}
B.~Boisseau, G.~Esposito-Farese, D.~Polarski, A.A. Starobinsky,
  Phys.~Rev.~Lett. \textbf{85}, 2236 (2000).
\newblock \doi{10.1103/PhysRevLett.85.2236}.
\newblock [\href{http://arxiv.org/abs/gr-qc/0001066}{{\tt gr-qc/0001066}}]

\bibitem{Gannouji:2006jm}
R.~Gannouji, D.~Polarski, A.~Ranquet, A.A. Starobinsky, JCAP \textbf{09}, 016
  (2006).
\newblock \doi{10.1088/1475-7516/2006/09/016}.
\newblock [\href{http://arxiv.org/abs/astro-ph/0606287}{{\tt
  astro-ph/0606287}}]

\bibitem{Carloni:2007eu}
S.~Carloni, S.~Capozziello, J.A. Leach, P.K.S. Dunsby, Class.~Quant.~Grav.
  \textbf{25}, 035008 (2008).
\newblock \doi{10.1088/0264-9381/25/3/035008}.
\newblock [\href{http://arxiv.org/abs/gr-qc/0701009}{{\tt gr-qc/0701009}}]

\bibitem{Bezrukov:2007ep}
F.L. Bezrukov, M.~Shaposhnikov, Phys.~Lett. \textbf{B659}, 703 (2008).
\newblock \doi{10.1016/j.physletb.2007.11.072}.
\newblock [\href{http://arxiv.org/abs/0710.3755}{{\tt arXiv:0710.3755}}]

\bibitem{Kamenshchik:1995ib}
A.Y. Kamenshchik, I.M. Khalatnikov, A.V. Toporensky, Phys.~Lett. \textbf{B357},
  36 (1995).
\newblock \doi{10.1016/0370-2693(95)00834-8}.
\newblock [\href{http://arxiv.org/abs/gr-qc/9508034}{{\tt gr-qc/9508034}}]

\bibitem{Hrycyna:2007gd}
O.~Hrycyna, M.~Szydlowski, Phys.~Rev. \textbf{D76}, 123510 (2007).
\newblock \doi{10.1103/PhysRevD.76.123510}.
\newblock [\href{http://arxiv.org/abs/0707.4471}{{\tt arXiv:0707.4471}}]

\bibitem{Hrycyna:2008gk}
O.~Hrycyna, M.~Szydlowski, JCAP \textbf{04}, 026 (2009).
\newblock \doi{10.1088/1475-7516/2009/04/026}.
\newblock [\href{http://arxiv.org/abs/0812.5096}{{\tt arXiv:0812.5096}}]

\bibitem{Hrycyna:2009zj}
O.~Hrycyna, M.~Szydlowski, Phys.~Lett. \textbf{B694}, 191 (2010).
\newblock \doi{10.1016/j.physletb.2010.09.061}.
\newblock [\href{http://arxiv.org/abs/0906.0335}{{\tt arXiv:0906.0335}}]

\bibitem{Hrycyna:2010yv}
O.~Hrycyna, M.~Szydlowski, JCAP \textbf{12}, 016 (2010).
\newblock \doi{10.1088/1475-7516/2010/12/016}.
\newblock [\href{http://arxiv.org/abs/1008.1432}{{\tt arXiv:1008.1432}}]

\bibitem{Hrycyna:2015eta}
O.~Hrycyna, M.~Szydlowski, JCAP \textbf{11}, 013 (2015).
\newblock \doi{10.1088/1475-7516/2015/11/013}.
\newblock [\href{http://arxiv.org/abs/1506.03429}{{\tt arXiv:1506.03429}}]

\bibitem{DeSimone:2008ei}
A.~De~Simone, M.P. Hertzberg, F.~Wilczek, Phys.~Lett. \textbf{B678}, 1 (2009).
\newblock \doi{10.1016/j.physletb.2009.05.054}.
\newblock [\href{http://arxiv.org/abs/0812.4946}{{\tt arXiv:0812.4946}}]

\bibitem{Bezrukov:2008ej}
F.L. Bezrukov, A.~Magnin, M.~Shaposhnikov, Phys.~Lett. \textbf{B675}, 88
  (2009).
\newblock \doi{10.1016/j.physletb.2009.03.035}.
\newblock [\href{http://arxiv.org/abs/0812.4950}{{\tt arXiv:0812.4950}}]

\bibitem{Barvinsky:2009fy}
A.O. Barvinsky, A.Y. Kamenshchik, C.~Kiefer, A.A. Starobinsky, C.~Steinwachs,
  JCAP \textbf{12}, 003 (2009).
\newblock \doi{10.1088/1475-7516/2009/12/003}.
\newblock [\href{http://arxiv.org/abs/0904.1698}{{\tt arXiv:0904.1698}}]

\bibitem{Clark:2009dc}
T.E. Clark, B.~Liu, S.T. Love, T.~ter Veldhuis, Phys.~Rev. \textbf{D80}, 075019
  (2009).
\newblock \doi{10.1103/PhysRevD.80.075019}.
\newblock [\href{http://arxiv.org/abs/0906.5595}{{\tt arXiv:0906.5595}}]

\bibitem{Belinskii:1985}
V.~Belinskii, L.~Grishchuk, Y.~Zel'dovich, I.~Khalatnikov, Sov.~Phys.~JETP
  \textbf{62}, 195 (1985).
\newblock [Zh.~Eksp.~Teor.~Fiz.~ {\bf 89} (1985) 346-360]

\bibitem{Belinsky:1985zd}
V.A. Belinsky, I.M. Khalatnikov, L.P. Grishchuk, Y.B. Zeldovich, Phys.~Lett.
  \textbf{B155}, 232 (1985).
\newblock \doi{10.1016/0370-2693(85)90644-6}

\bibitem{Belinskii:1987}
V.~Belinskii, I.~Khalatnikov, Sov.~Phys.~JETP \textbf{66}, 441 (1987).
\newblock [Zh.~Eksp.~Teor.~Fiz. {\bf 93} (1987) 785-799]

\bibitem{Wainwright:book}
J.~Wainwright, G.F.R. Ellis (eds.), \emph{{Dynamical Systems in Cosmology}}
  (Cambridge University Press, Cambridge, 1997)

\bibitem{Humieja:2019ywy}
F.~Humieja, M.~Szydlowski, Eur.~Phys.~J. \textbf{C79}(9), 794 (2019).
\newblock \doi{10.1140/epjc/s10052-019-7299-x}.
\newblock [\href{http://arxiv.org/abs/1901.06578}{{\tt arXiv:1901.06578}}]

\bibitem{Faraoni:1999hp}
V.~Faraoni, E.~Gunzig, Int.~J.~Theor.~Phys. \textbf{38}, 217 (1999).
\newblock \doi{10.1023/A:1026645510351}.
\newblock [\href{http://arxiv.org/abs/astro-ph/9910176}{{\tt
  astro-ph/9910176}}]

\bibitem{Kamenshchik:2014waa}
A.{\relax Yu}. Kamenshchik, C.F. Steinwachs, Phys.~Rev. \textbf{D91}, 084033
  (2015).
\newblock \doi{10.1103/PhysRevD.91.084033}.
\newblock [\href{http://arxiv.org/abs/1408.5769}{{\tt arXiv:1408.5769}}]

\bibitem{Bahamonde:2017kbs}
S.~Bahamonde, S.D. Odintsov, V.K. Oikonomou, P.V. Tretyakov, Phys.~Lett.
  \textbf{B766}, 225 (2017).
\newblock \doi{10.1016/j.physletb.2017.01.012}.
\newblock [\href{http://arxiv.org/abs/1701.02381}{{\tt arXiv:1701.02381}}]

\bibitem{Calmet:2017voc}
X.~Calmet, I.~Kuntz, Eur.~Phys.~J. \textbf{C77}(2), 132 (2017).
\newblock \doi{10.1140/epjc/s10052-017-4695-y}.
\newblock [\href{http://arxiv.org/abs/1702.03832}{{\tt arXiv:1702.03832}}]

\bibitem{Planck2018:X}
{\bf Planck} Collaboration, Y.~Akrami, et~al.,
  [\href{http://arxiv.org/abs/1807.06211}{{\tt arXiv:1807.06211}}]

\bibitem{Alho:2015cza}
A.~Alho, J.~Hell, C.~Uggla, Class.~Quant.~Grav. \textbf{32}, 145005 (2015).
\newblock \doi{10.1088/0264-9381/32/14/145005}.
\newblock [\href{http://arxiv.org/abs/1503.06994}{{\tt arXiv:1503.06994}}]

\bibitem{Perko:book}
L.~Perko, \emph{Differential Equations and Dynamical Systems}, \emph{Texts in
  Applied Mathematics}, vol.~7, 3rd edn. (Springer-Verlag, New York, 2001)

\bibitem{Wiggins:book}
S.~Wiggins, \emph{Introduction to Applied Nonlinear Dynamical Systems and
  Chaos}, \emph{Texts in Applied Mathematics}, vol.~2, 2nd edn.
  (Springer-Verlag, New York, 2003)

\bibitem{Dutta:2020uha}
J.~Dutta, L.~Jarv, W.~Khyllep, S.~Tokke,   (2020).
\newblock [\href{http://arxiv.org/abs/2007.06601}{{\tt arXiv:2007.06601}}]

\bibitem{Linde:2001ae}
A.D. Linde, JHEP \textbf{11}, 052 (2001).
\newblock \doi{10.1088/1126-6708/2001/11/052}.
\newblock [\href{http://arxiv.org/abs/hep-th/0110195}{{\tt hep-th/0110195}}]

\bibitem{Kofman:2007tr}
L.~Kofman, S.~Mukohyama, Phys. Rev. \textbf{D77}, 043519 (2008).
\newblock \doi{10.1103/PhysRevD.77.043519}.
\newblock [\href{http://arxiv.org/abs/0709.1952}{{\tt arXiv:0709.1952}}]

\bibitem{Chiba:2008ia}
T.~Chiba, M.~Yamaguchi, JCAP \textbf{10}, 021 (2008).
\newblock \doi{10.1088/1475-7516/2008/10/021}.
\newblock [\href{http://arxiv.org/abs/0807.4965}{{\tt arXiv:0807.4965}}]

\bibitem{Felder:2002jk}
G.N. Felder, A.V. Frolov, L.~Kofman, A.D. Linde, Phys. Rev. D \textbf{66},
  023507 (2002).
\newblock \doi{10.1103/PhysRevD.66.023507}.
\newblock [\href{http://arxiv.org/abs/hep-th/0202017}{{\tt hep-th/0202017}}]

\bibitem{Boisseau:2015hqa}
B.~Boisseau, H.~Giacomini, D.~Polarski, A.A. Starobinsky, JCAP \textbf{07}, 002
  (2015).
\newblock \doi{10.1088/1475-7516/2015/07/002}.
\newblock [\href{http://arxiv.org/abs/1504.07927}{{\tt arXiv:1504.07927}}]

\bibitem{Starobinsky:1980te}
A.A. Starobinsky, Phys.~Lett. \textbf{B91}, 99 (1980).
\newblock \doi{10.1016/0370-2693(80)90670-X}

\bibitem{Mukhanov:1991zn}
V.F. Mukhanov, R.H. Brandenberger, Phys.~Rev.~Lett. \textbf{68}, 1969 (1992).
\newblock \doi{10.1103/PhysRevLett.68.1969}

\bibitem{Brandenberger:1993ef}
R.H. Brandenberger, V.F. Mukhanov, A.~Sornborger, Phys.~Rev. \textbf{D48}, 1629
  (1993).
\newblock \doi{10.1103/PhysRevD.48.1629}.
\newblock [\href{http://arxiv.org/abs/gr-qc/9303001}{{\tt gr-qc/9303001}}]

\bibitem{tHooft:2014daa}
G.~'t~Hooft, Int.~J.~Mod.~Phys.~D \textbf{24}, 1543001 (2015).
\newblock \doi{10.1142/S0218271815430014}.
\newblock [\href{http://arxiv.org/abs/1410.6675}{{\tt arXiv:1410.6675}}]

\bibitem{Hrycyna:2017oug}
O.~Hrycyna, Acta Phys. Polon. Supp. \textbf{10}, 425 (2017).
\newblock \doi{10.5506/APhysPolBSupp.10.425}.
\newblock [\href{http://arxiv.org/abs/1705.10593}{{\tt arXiv:1705.10593}}]

\end{thebibliography}

\end{document}